\documentclass[prc,preprint,showpacs,showkeys,
        tightenlines,amsfonts,nofootinbib]{revtex4}

\usepackage{graphicx}  %
\usepackage{bm}  %
\usepackage{amssymb}
\usepackage{dcolumn}

\begin{document}

%


\newcommand{\beq}{\begin{equation}}
\newcommand{\eeq}{\end{equation}}
\newcommand{\bea}{\begin{eqnarray}}
\newcommand{\eea}{\end{eqnarray}}
\newcommand{\ben}{\begin{eqnarray*}}
\newcommand{\een}{\end{eqnarray*}}

\newcommand{\simlt}{\stackrel{<}{{}_\sim}}
\newcommand{\simgt}{\stackrel{>}{{}_\sim}}
\newcommand{\sing}{$^1\!S_0$ }
\newcommand{\btau}{\mbox{\boldmath$\tau$}}
\newcommand{\bsig}{\mbox{\boldmath$\sigma$}}

\newcommand{\dt}{\partial_t}

\newcommand{\kf}{k_{\rm F}}
\newcommand{\wt}{\widetilde}
\newcommand{\kt}{\widetilde k}
\newcommand{\pt}{\widetilde p}
\newcommand{\qt}{\widetilde q}
\newcommand{\wh}{\widehat}
\newcommand{\dens}{\rho}
\newcommand{\edens}{{\cal E}}
\newcommand{\order}[1]{{\cal O}(#1)}

\newcommand{\psihat}{\widehat\psi}
\newcommand{\xvec}{{\bf x}}
\newcommand{\dagphan}{{\phantom{\dagger}}}
\newcommand{\kvec}{{\bf k}}
\newcommand{\kpvec}{{\bf k}'}
\newcommand{\ak}{a^\dagphan_\kvec}
\newcommand{\akdag}{a^\dagger_\kvec}
\newcommand{\akv}[1]{a^\dagphan_{\kvec{#1}}}
\newcommand{\akdagv}[1]{a^\dagger_{\kvec{#1}}}
\newcommand{\amkv}[1]{a^\dagphan_{-\kvec{#1}}}
\newcommand{\amkdagv}[1]{a^\dagger_{-\kvec{#1}}}
\newcommand{\akp}{a^\dagphan_{\kvec'}}
\newcommand{\akpdag}{a^\dagger_{\kvec'}}
\newcommand{\akpv}[1]{a^\dagphan_{\kvec'{#1}}}
\newcommand{\akpdagv}[1]{a^\dagger_{\kvec'{#1}}}
\newcommand{\alphak}{\alpha_{\kvec}}
\newcommand{\alphakp}{\alpha_{\kvec'}}
\newcommand{\alphamk}{\alpha_{-\kvec}}
\newcommand{\betak}{\beta_{\kvec}}
\newcommand{\betakp}{\beta_{\kvec'}}
\newcommand{\betamk}{\beta_{-\kvec}}

\def\vec#1{{\bf #1}}

\newcommand{\nab}{\overrightarrow{\nabla}}
\newcommand{\nabsq}{\overrightarrow{\nabla}^{2}\!}
\newcommand{\nabl}{\overleftarrow{\nabla}}
\newcommand{\galnab}{\tensor{\nabla}}
\newcommand{\psid}{{\psi^\dagger}}
\newcommand{\psidal}{{\psi^\dagger_\alpha}}
\newcommand{\psidbe}{{\psi^\dagger_\beta}}
\newcommand{\idt}{{i\partial_t}}
\newcommand{\Sthree}{{\delta_{11'}(\delta_{22'}\delta_{33'}%
        -\delta_{23'}\delta_{32'})%
        +\delta_{12'}(\delta_{23'}\delta_{31'}-\delta_{21'}\delta_{33'})%
        +\delta_{13'}(\delta_{21'}\delta_{32'}-\delta_{22'}\delta_{31'})}}
\newcommand{\Stwo}{{\delta_{11'}\delta_{22'}-\delta_{12'}\delta_{21'}}}
\newcommand{\Left}{{\cal L}}
\newcommand{\Leuclid}{{\cal L}_{E}}
\newcommand{\Tr}{{\rm Tr}}

\newcommand{\h}{\hfil}
\newcommand{\be}{\begin{enumerate}}
\newcommand{\ee}{\end{enumerate}}
\newcommand{\I}{\item}   

\newcommand{\density}{\rho}

\newcommand{\thyp}{\mbox{---}}

\newcommand{\Jks}{J_{\ks}}
\newcommand{\Jzero}{\Jks}
\newcommand{\Jdensityzero}{J_\density^0}

\newcommand{\ks}{{\rm ks}}
\newcommand{\Seq}{Schr\"odinger\ equation }
\newcommand{\yvec}{{\bf y}}
\newcommand{\ve}{V_{eff}}
\newcommand{\densityJzero}{\density^0_J}
\newcommand{\Dfunct}{{D^{-1}}}
\newcommand{\drv}[2]{{\mbox{$\partial$} #1\over \mbox{$\partial$} #2}}
\newcommand{\drvs}[2]{{\mbox{$\partial^2$} #1\over \mbox{$\partial$} #2 \mbox{$^2$}}}
\newcommand{\drvt}[2]{{\partial^3 #1\over \partial #2 ^3}}
\newcommand{\til}[1]{{\widetilde #1}}
\newcommand{\dthreex}{d^3\xvec}
\newcommand{\dthreey}{d^3\yvec}

\newcommand{\efermi}{\varepsilon_{{\scriptscriptstyle \rm F}}}
\newcommand{\eHF}{\wt\varepsilon}
\newcommand{\eKS}{e}
\newcommand{\ekJ}{e_\kvec^J}
\newcommand{\epsk}{\varepsilon_\kvec}
\newcommand{\epsKS}{\varepsilon}
\newcommand{\Eq}[1]{Eq.~(\ref{#1})}

\newcommand{\Fi}[1]{\mbox{$F_{#1}$}}
\newcommand{\fq}{f_{\qvec}}
\newcommand{\FKS}{F_{\ks}}

\newcommand{\Gammaalt}{\overline\Gamma}
\newcommand{\Gammaks}{\Gamma_{\ks}}
\newcommand{\Gamint}{\widetilde{\Gamma}_{\rm int}}
\newcommand{\GKS}{G_{\ks}}
\newcommand{\GKSmatrix}{{\bf G}_{\ks}}
\newcommand{\grad}{{\bm{\nabla}}}   
\newcommand{\greenKS}{{G}_{\ks}}
\newcommand{\intint}{\int\!\!\int}

\newcommand{\Gmatrix}{{\bf G}_0}
\newcommand{\Geucl}{{\cal G}_0}
\newcommand{\Feucl}{{\cal F}_0}

\newcommand{\kfermi}{k_{{\scriptscriptstyle \rm F}}}   
\newcommand{\kJzero}{k_J}

\renewcommand{\l}{\lambda}

\newcommand{\MeV}{\mbox{\,MeV}}
\newcommand{\mi}[1]{\mbox{$\mu_{#1}$}}

\newcommand{\Oi}[1]{\mbox{$\Omega_{#1}$}}

\newcommand{\phibar}{\overline\phi}
\newcommand{\phidagger}{\phi^\dagger}
\newcommand{\phistar}{\phi^\ast}
\newcommand{\psibar}{\overline\psi}
\newcommand{\psidagger}{\psi^\dagger}
\newcommand{\qvec}{\vector{q}}

\newcommand{\tr}{{\rm tr\,}}

\newcommand{\Ulong}{U_{L}}

\renewcommand{\vector}[1]{{\bf #1}}
\newcommand{\vext}{v_{\rm ext}}   
\newcommand{\Vlong}{V_{L}}

\newcommand{\Wzero}{W_0}
\newcommand{\Wks}{W_{\ks}}
\newcommand{\zvec}{\vector{z}}
\newcommand{\rvec}{\vector{r}}

\newcommand{\psiup}{\psi_{\uparrow}}
\newcommand{\psidagup}{\psi^\dagger_{\uparrow}}
\newcommand{\psidown}{\psi_{\downarrow}}
\newcommand{\psidagdown}{\psi^\dagger_{\downarrow}}

\newcommand{\gs}{{\rm gs}}
\newcommand{\ts}{\textstyle}

\newcommand{\hzero}{\widehat h_0}

\newcommand{\etadag}{\eta^{\dagger}}
\newcommand{\BCSbra}{\langle {\rm BCS} |}
\newcommand{\BCSket}{| {\rm BCS} \rangle}

\newcommand{\jpair}{j}
\newcommand{\jzpair}{j_0}
\newcommand{\Hhat}{{\widehat H}}

\newcommand{\phibare}{\phi_{\rm B}}
%


\title{Effective Field Theory for \\
        Dilute Fermions with Pairing}

\author{R.J. Furnstahl}\email{furnstahl.1@osu.edu}
\affiliation{Department of Physics,
         The Ohio State University, Columbus, OH\ 43210}
\author{H.-W.\ Hammer}\email{hammer@itkp.uni-bonn.de}
\affiliation{Helmholtz-Institut f\"ur Strahlen- und Kernphysik (Theorie),
Universit\"at Bonn, Nu\ss allee 14-16, D-53115 Bonn, Germany}
\author{S.J.\ Puglia}
\email{spuglia@sbiguk.com}
\affiliation{SBIG (UK) PLC, Berkeley Square House, London W1J 6BR,
United Kingdom}

%
\date{\today}
%

\begin{abstract}
Effective field theory (EFT) methods for a 
uniform system of fermions with short-range, natural interactions 
are extended to include pairing correlations,
as part of a program to develop a systematic
Kohn-Sham density functional theory (DFT)
for medium and heavy nuclei.
An effective action formalism for local composite operators
leads to a free-energy functional that includes pairing
by applying an inversion method order by order in the EFT expansion.
A consistent renormalization scheme is demonstrated for the uniform system 
through next-to-leading order, which includes
induced-interaction corrections to pairing.
\end{abstract}

\smallskip
\maketitle


\section{Introduction}
\label{sec:introduction}

In previous work, effective field theory (EFT)
\cite{LEPAGE89,BEANE99,Bedaque:2002mn,Kaplan:2005es,Epelbaum:2005pn,Braaten:2004rn}
was applied to the normal ground state
of a dilute Fermi system with short-range
interactions  \cite{HAMMER00,Furnstahl:2002gt} and subsequently
extended to finite systems as a realization of Kohn-Sham density
functional theory (DFT) \cite{PUGLIA03,FURNSTAHL04,FURNSTAHL04b}. 
These studies are part of a program to go beyond mean-field approaches
to finite many-body systems
while maintaining their computational simplicity and phenomenological
strengths.
Our ultimate goal is to systematically describe
medium and heavy nuclei using EFT and renormalization group methods, 
including nuclei far from
stability where EFT power counting and error estimates are desirable
to enable reliable extrapolations. 
For nuclei as well as trapped systems of atomic gases,
attractive interactions cause  
the normal ground state to be unstable to the formation of Cooper pairs.
In the present work, we consider an extension of the EFT formalism
for uniform systems
to allow for pairing correlations in a way that is also consistent
with Kohn-Sham DFT. 

Our strategy is to 
apply an effective action formalism \cite{COLEMAN88,PESKIN95,WEINBERG96}
to calculate the free energy order by order in an EFT expansion.  
The effective action formalism is often used in condensed matter
applications to discuss superconductivity, starting from a path
integral with a four-fermion contact interaction (see, for example,
\cite{NAGAOSA,STONE}).
In the standard approach, a pairing field is introduced as a 
charged scalar auxiliary
field and shifted to eliminate the four-fermion term, leaving only
a Gaussian integral over the fermion fields.
After performing this integral, a conventional effective
action is derived by a functional
Legendre transformation with respect to the
auxiliary field.
A minimum at a nonzero expectation value of this field (which is
proportional to the pairing gap for a short-range interaction at leading
order) indicates spontaneous symmetry
breaking of the phase symmetry related to fermion number conservation,
which implies the normal ground state is unstable to pairing and
determines the ``superconducting'' ground state.

Instead of introducing an auxiliary pairing field, we couple an external source 
(or, more generally, a pair of
complex conjugate sources) to the local,
composite pair density [Eq.~(\ref{partfunca})].
In general this source would be space-time dependent, but for a uniform system 
in its ground state, which we consider here, the source 
can be treated as a constant.
This external source
is analogous to the source introduced for finite systems
in Ref.~\cite{PUGLIA03},
which was coupled to the composite fermion density.
As in that case, we carry out a functional Legendre transformation
order by order in the EFT expansion via the inversion method introduced by
Fukuda et al.\ \cite{FUKUDA94,FUKUDA95}
(see Ref.~\cite{INAGAKI92} for an application to superconductivity).
If we keep the space dependence of the sources and the corresponding
densities, we obtain a generalization of the EFT-based density
functional theory (DFT) formalism of Ref.~\cite{PUGLIA03}
(see also
Refs.~\cite{VALIEV97,VALIEV97b,VALIEV96,CHITRA00,CHITRA01,Polonyi:2001uc}).
In the present work, the limitation to uniform systems means that
the inversion method parallels the construction used long ago by
Kohn, Luttinger, and Ward \cite{KOHN60,LUTTINGER60} to show the
relationship of zero-temperature diagrammatic calculations to ones
using the finite-temperature Matsubara formalism in the zero-temperature
limit.

In Refs.~\cite{OLIVEIRA88} and \cite{KURTH99}, density functional theory for
superconductivity was proposed using a nonlocal source coupled to the nonlocal
pair density.
While this approach can be embedded in an effective action framework,
we first want to explore the use of local sources, which are greatly preferred
for finite nuclei from a computational point of view.
In particular, we seek 
a consistent generalization of
the Skyrme-Hartree-Fock-Bogoliubov approach \cite{Stoitsov:2003pd}.

Potential problems arise, however, because
the use of  local composite
operators and zero-range interactions leads to new ultraviolet
divergences even at the
mean-field (Hartree-Fock) level when pairing is included.
Divergences at this order are not encountered when coupling an external
source to the fermion density $\psi^\dagger\psi$, but appear now because   
the composite operators
$\psi\psi$ and $\psi^\dagger\psi^\dagger$  need additional 
renormalization \cite{COLLINS86}.
The divergences at leading order
are symptomatic of generic 
problems identified long ago by 
Banks and Raby \cite{BANKS76} that arise with effective 
potentials of local composite operators.
(Such divergences would be avoided by working with nonlocal sources
as in Refs.~\cite{OLIVEIRA88,KURTH99}.)
These problems inhibited for many years
the use of effective actions of composite
operators although more
recently Verschelde et al.\ \cite{VERSCHELDE95,VERSCHELDE97,KNECT01}
and Miransky et al.\ \cite{MIRANSKY93,MIRANSKY97,MIRANSKY98} have revived
their use for relativistic field theories.
In this work, we show how to 
remove the new UV divergences through next-to-leading-order (NLO) 
by adding a term quadratic in the pairing source,
using a renormalization prescription compatible with the inversion
method and conventional treatments of pairing at leading order.

The
BCS gap equation for contact interactions
has conventionally been renormalized at leading order (LO)
by using the equation for
free-space scattering, which has the same divergence [see
Eqs.~(\ref{eq:gap1}) and (\ref{eq:freespace})].
The four-fermion coupling is eliminated in favor of the scattering length,
leaving finite, renormalized equations.
This approach was applied in Refs.~\cite{MARINI98} and \cite{PaB99}
to obtain analytic formulas for the pairing gap and ground state energy at
LO.
In Ref.~\cite{PaB99}, dimensional regularization with minimal subtraction
(DR/MS) was used.  
DR/MS
was also used to good advantage for the dilute Fermi calculations 
of the normal ground state in
Ref.~\cite{HAMMER00}.  
In extending the formalism of Ref.~\cite{HAMMER00} 
to include
pairing, we use a more general
subtraction scheme (called ``power divergence subtraction,'' or PDS),
that lets us 
verify explicitly that the generating functional, effective action,
gap equation, and observables are renormalized order by order in the inversion
approach. 
We give renormalized expressions up to
next-to-leading order (NLO), which contain the universal
correction to the pre-factor of the gap at low density, first
calculated in Ref.~\cite{GORKOV61} (see also Ref.~\cite{HEISELBERG00}).

The plan of the paper is as follows.  
In Sect.~\ref{sect:klw}, we briefly rederive the EFT for a dilute Fermi gas
in the effective action formalism using the inversion method
of
Kohn, Luttinger, and Ward \cite{KOHN60}.
In Sect.~\ref{sect:effact}, the approach is generalized to include
pairing by introducing a source coupled to the pairing density and
then carrying out the 
Legendre transformation to get an effective action as a functional
of fermion and pair densities.
The construction of the EFT effective action functional via the
inversion method starts
with the zeroth order functional.  
In Sect.~\ref{sect:Gammaone}, we carry out the leading order (LO)
calculation in detail, reproducing standard results and extending the
dimensional regularization results of Ref.~\cite{PaB99}.
In Sect.~\ref{sect:Gammatwo}, we extend the calculation to
next-to-leading order (NLO), which introduces the so-called
``induced interaction''.
Section~\ref{sect:summary} is a summary and
some additional details are included in the Appendices.


\section{Inversion Method Without Pairing}
\label{sect:klw}

\subsection{Lagrangian and Generating Functional}
\label{dil1a}

In contrast to the zero-temperature,
Minkowski-space EFT treatment of dilute systems in
Ref.~\cite{HAMMER00}, we work in Euclidean space in the Matsubara
(imaginary time) formalism at inverse temperature $\beta$.  
We use a four-vector notation $x \equiv \{\tau,\xvec\}$, with the
$\tau$ dependence usually implicit, or $\kt \equiv \{\omega_n,\kvec\}$
in momentum space, where $\omega_n$ is a fermion Matsubara frequency.
Throughout this work, however, we will consider the zero temperature 
($\beta\rightarrow\infty$) limit, for which the differences from
Ref.~\cite{HAMMER00} are minor (and the results equivalent after
frequency integrals are performed).

The EFT Lagrangian
for a nonrelativistic spin-1/2 fermion
field with spin-independent interactions
is the same as in Ref.~\cite{HAMMER00}, but converted to Euclidean
form (which means $t \rightarrow -i\tau$ and an overall minus sign
in defining $\Leuclid$ and the interaction Lagrangian
$\Leuclid^{\rm int}$):
\bea
  \Leuclid  &=&
       \psi^\dagger \biggl[\frac{\partial}{\partial\tau}  
          - \frac{\nab^{\,2}}{2M}\biggr]
                 \psi + \frac{C_0}{2}(\psi^\dagger \psi)^2
            - \frac{C_2}{16}\Bigl[ (\psi\psi)^\dagger
                                  (\psi\galnab^2\psi)+\mbox{ h.c.}
                             \Bigr]
  \nonumber \\[5pt]
   & & \null -
         \frac{C_2'}{8} (\psi \galnab \psi)^\dagger \cdot
             (\psi\galnab \psi)
+  \ldots
  \nonumber \\
   &\equiv& 
    \psi^\dagger \biggl[\frac{\partial}{\partial\tau}  
          - \frac{\nab^{\,2}}{2M}\biggr] \psi
	  + \Leuclid^{\rm int}(\psi^\dagger,\psi) \ ,
  \label{eq:lag}                                                   
\eea
where $\galnab=\overleftarrow{\nabla}-\nab$ is the Galilean invariant
derivative and h.c.\ denotes the Hermitian conjugate.%
\footnote{We use units with $\hbar = 1$.}
The terms proportional to $C_2$ and $C_2'$ contribute to $s$-wave and
$p$-wave scattering, respectively, 
while the dots represent terms with more derivatives and/or more
fields, as well as renormalization counterterms.
The Lagrangian Eq.~(\ref{eq:lag}) is a particular 
canonical form, which can be reached via field redefinitions.
For example,
higher-order terms with time derivatives are omitted, as they can be
eliminated in favor of terms with spatial derivatives (see, for example,
Ref.~\cite{Furnstahl:2000we}).
We will restrict ourselves here to spin-1/2 (i.e., degeneracy $\nu=2$)
and spin-independent interactions,
which is sufficient to illustrate the formalism and the renormalization
issues; generalizing to isospin and spin-dependent interations
is straightforward. 
We also consider only the $C_0$ vertex here, but comment at the end
on the inclusion of vertices with derivatives.

We first consider a Euclidean generating functional with chemical
potential $\mu$ and external sources
$\eta(x)$ and $\etadag(x)$ \cite{NEGELE88,STONE}: 
\beq
  Z[\eta,\etadag; \mu] \equiv e^{-W[\eta,\etadag; \mu]}
      = \int\! D\psi_\alpha D\psi_\alpha^\dag 
      \ e^{-\int\! d^4x\, [\Leuclid\,-\,   \mu\,
    \psi_\alpha^\dag(x) \psi_\alpha(x)
    + \etadag_\alpha(x)\psi_\alpha(x) +
    \psi_\alpha^\dag(x)\eta_\alpha(x)]}
    \ ,
   \label{partfunc1}
\eeq
where $\int d^4x$ includes a $d\tau$ integration that
runs from $-\beta/2$ to $\beta/2$ (to facilitate the
$\beta\rightarrow\infty$ limit) and
anti-periodic boundary conditions are imposed.
For simplicity, normalization factors are considered to be implicit
in the functional integration measure.
(See Refs.~\cite{NEGELE88,FUKUDA94,FUKUDA95} for more detailed
treatments of similar path integrals.)
We have written the spin index $\alpha$ 
explicitly in Eq.~(\ref{partfunc1}); we
will interchangeably use $\alpha = \{1,2\}$ and 
$\alpha = \{\uparrow,\downarrow\}$ in the sequel to denote individual
spins, and
the spin indices will be left implicit where there is no chance of confusion.
We have kept the chemical potential $\mu$ separate from the Lagrangian in
Eq.~(\ref{partfunc1}) to emphasize its role as an external
source.  In the next section, we will add a corresponding source coupled
to the pair density.

\begin{figure}[t]
\begin{center}
  \includegraphics*[width=12.cm,angle=0]{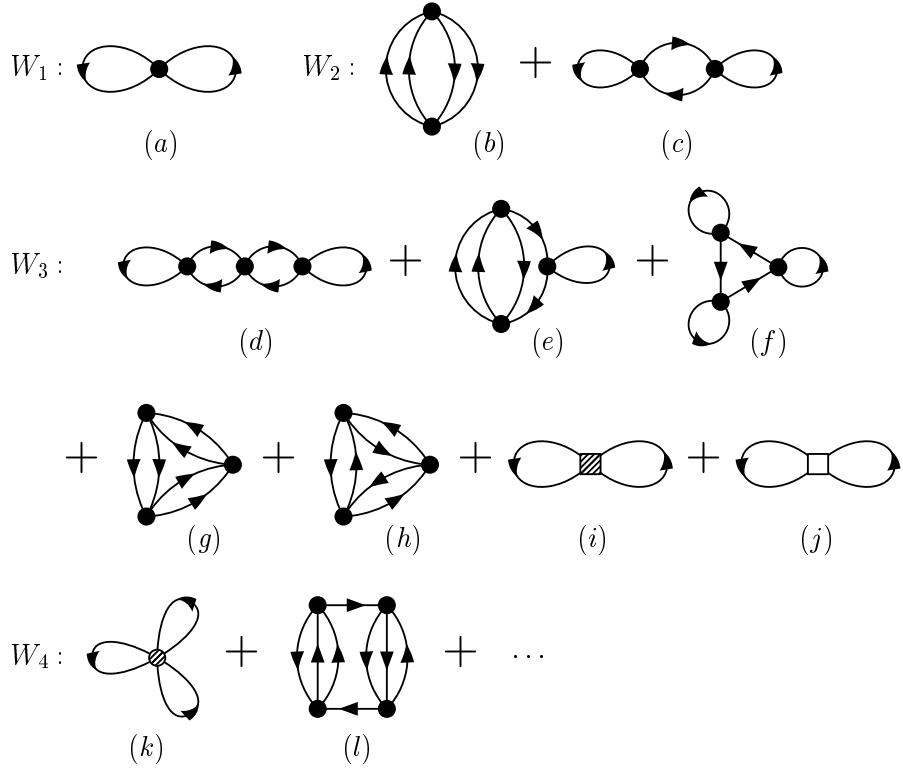}
\end{center}
\caption{Hugenholtz diagrams for the unrenormalized $W_i$ functionals
in a homogeneous, dilute Fermi system.  The vertices are defined
in Fig.~\ref{eftvertex}.}
\label{fig_Wexpansion}
\end{figure}

A conventional perturbative expansion is realized by removing the
interaction terms from the path integral in (\ref{eq:lag}) in favor of
functional derivatives with respect to $\eta$ and $\etadag$ and
performing the remaining Gaussian integration over $\psi$
and $\psi^\dagger$ \cite{NEGELE88,STONE}:
\beq
  Z[\eta,\etadag; \mu] = Z_0\,
    e^{-\int\! d^4x\, \Leuclid^{\rm int}[\delta/\delta\eta(x),
      -\delta/\delta\eta^\dagger(x)]}
      \,
    e^{\int\! d^4y\, d^4y'\, \eta^\dagger(y)
    \Geucl(y,y') \eta(y')} \ ,       
\eeq
where the spin indices are implicit and we have introduced the
noninteracting partition function $Z_0$.
Explicit expressions for 
the Matusbara Green's function $\Geucl$ in coordinate and momentum
space can be found in Ref.~\cite{NEGELE88}.
The linked-cluster theorem \cite{NEGELE88} shows that the difference of
the interacting and noninteracting thermodynamic
potentials $\Omega$ and $\Omega_0$ is given by 
the sum of connected diagrams from the expansion of $Z$, with
the external sources $\eta^\dagger$ and $\eta$ set to zero at the end:
\beq
  \Omega(\mu,T,V) - \Omega_0(\mu,T,V) =  
     \frac{1}{\beta}(W[0,0;\mu] - W_0[0,0;\mu]) \ .
     \label{eq:Omega}
\eeq
The connected Feynman diagrams that sum (with appropriate symmetry factors)
to $W-W_0$, which are labled $W_i$ with $i\geq 1$,
are shown in Fig.~\ref{fig_Wexpansion}, with the fermion
lines representing Matsubara propagators $\Geucl$ 
with chemical potential $\mu$ \cite{NEGELE88}.
The subscript $i$ labels the leading order
to which $W_i$ contributes in the dilute EFT expansion.
(In the DR/MS scheme, each diagram contributes to only one order but
this is no longer true in DR/PDS.)
The chemical potential can be determined for given $N$
by inverting the thermodynamic
expression for the particle number, $N = \partial\Omega/\partial\mu$.
The regularization and renormalization of divergences arising in the
evaluation of the $W_i$ are described below.

\subsection{Kohn-Luttinger-Ward Inversion}
\label{subsect:subsec}

The Kohn-Luttinger-Ward (KLW) theorem \cite{KOHN60,LUTTINGER60} relates
the perturbative calculation of diagrams using the finite-temperature
Matsubara formalism in the zero-temperature limit
to the calculation of
diagrams using zero-temperature perturbation theory, which was
applied to the dilute Fermi gas in Ref.~\cite{HAMMER00}.
In particular, the diagrammatic expansion
of $\Omega(\mu,V,T\rightarrow 0)$ is, under specified conditions, 
the same as the free-energy $F(N,V,T\equiv 0)$ evaluated using
zero-temperature propagators with the non-interacting chemical
potential $\mu_0$ and with anomalous diagrams omitted
(see Ref.~\cite{HAMMER00} for Feynman rules).
The anomalous diagrams are those such as Fig.~\ref{fig_Wexpansion}(c)
and (d), which vanish in zero-temperature perturbation
theory because they have particles and holes in the same intermediate
single-particle state (i.e., factors of
$\theta(\mu_0-\varepsilon_\alpha)\theta(\varepsilon_\alpha-\mu_0)=0$ appear).
The conditions of the theorem
require that the symmetry of the non-interacting
and interacting systems agree \cite{FETTER71,NEGELE88}.

A demonstration of the KLW theorem using an inversion method
for the case of an electron gas is presented
in Ref.~\cite{FETTER71}. 
We adapt this demonstration to the case of an EFT with a well-defined
expansion.  
While we illustrate the procedure with the expansion relevant to a natural
short-distance interaction (which is an expansion in the Fermi momentum
$\kf$ times the effective range parameters $a_s$, $r_s$, and so on
\cite{HAMMER00}), the
discussion is more general, requiring only a hierarchical expansion. 
Thus, for example, 
a nonperturbative (in diagrams) $1/N$ expansion works just as well.

The basic plan is to carry out the conventional thermodynamic
Legendre transformation:
\beq
   F(N) = \Omega(\mu) + \mu N \ ,
   \label{eq:LT1}
\eeq
with $\mu(N)$ obtained by inverting 
$N(\mu)=-(\partial\Omega/\partial\mu)_{TV}$,
which determines the mean-number $N$ of particles.
We expand each of the quantities in Eq.~(\ref{eq:LT1}) about the
non-interacting system:
\bea
       \Omega(\mu) &=& \Omega_0(\mu) + \Omega_1(\mu) + \Omega_2(\mu) +
        \cdots \ , \label{eq:Omegaexp} \\
        \mu &=& \mu_0 + \mu_1 + \mu_2 + \cdots \ , 
           \label{eq:muexp} \\
        F(N) &=& F_0(N) + F_1(N) + F_2(N) + \cdots \ ,
   \label{eq:Fexp}
\eea
where the subscript indicates the order of the expansion.
(Note that the subscript is just a counting parameter that
does not have to correspond to a \emph{power series} in the expansion
parameter; e.g., in the Coulomb case the
expansion parameter is $e^2$ but $\Omega_2$ has both an $e^4$ term and the
correlation energy of order $e^4\ln e$.)
The non-interacting system refers to a system of zeroth order in the EFT
expansion parameter.  This means the zeroth-order 
system has no \emph{internal} interactions among the particles, but it
can include external sources  (we exploit this freedom below).
Each $\Omega_i(\mu)$ is given by
$(1/\beta) W_i[0,0;\mu]$, where the $W_i$'s correspond to the diagrams in
Fig.~\ref{fig_Wexpansion}.

The number of particles is
\beq
  N(\mu,T,V) = -\left(\frac{\partial\Omega}{\partial\mu}\right)_{TV}
    = - \frac{\partial\Omega_0}{\partial\mu}
    - \frac{\partial\Omega_1}{\partial\mu}
    - \frac{\partial\Omega_2}{\partial\mu}
    + \cdots
    \label{eq:Nexp}
\eeq
Note that we could simply use the \emph{unexpanded} first equality in
Eq.~(\ref{eq:Nexp}) together with the series in Eq.~(\ref{eq:Omegaexp}),
since they define a parametric relation between $N$ and $\Omega$ in
terms of $\mu$ \cite{FETTER71}.
Since we have in mind the extension to finite systems, we pursue an
alternative that generalizes to Kohn-Sham density functional theory
(DFT).
To carry this out, 
we perform the inversion in Eq.~(\ref{eq:Nexp}) order by order
in the EFT expansion, treating $N$ as
order zero in the expansion.
(That is, we ensure there are no corrections to $N$ at higher order.)
This means that the zeroth order equation,
\beq
   N = - \left[ \frac{\partial\Omega_0}{\partial\mu} \right]_{\mu=\mu_0}
   \ ,
   \label{eq:Neq}
\eeq
is the \emph{only} equation to which $N$ contributes (by construction).  
Thus the
``exact'' $N$ is obtained from the non-interacting thermodynamic
potential.  This might not sound impressive, but the analogous situation
holds when we generalize $\mu$ to be position dependent  
or coupled to the pair density in a finite system.
In these cases, it is the exact, 
spatially dependent fermion or pair density 
(with appropriate renormalization conditions, see below)
that is obtained
from a non-interacting system with a single-particle potential that
is the generalization of $\mu_0$.
This is precisely the description of
the Kohn-Sham system (see Refs.~\cite{PUGLIA03} and \cite{FURNSTAHL04} 
for details on carrying
out the DFT case without pairing). 

Equation (\ref{eq:Neq}) determines $N(\mu_0)$ at any temperature, from
which we can find $\mu_0(N)$ 
for any
system for which we can identify $\Omega_0$
(the inversion is unique since $\mu_0$ is a monotonic function of $N$
\cite{FETTER71}).
If we have a uniform system with no external sources,
$\mu_0$ is the chemical potential of a noninteracting 
Fermi gas at temperature $T$ with density $N/V$.  
In particular, at $T=0$
with no external potential and spin-isospin
degeneracy $\nu$,
 \beq
    \mu_0(N) = (6\pi^2 N/\nu V)^{2/3} \equiv \kf^2/2M \equiv
            \efermi^0 \ .
       \label{eq:mu0}
 \eeq
The first-order equation extracted from Eq.~(\ref{eq:Nexp})
has two terms, which lets us solve for
$\mu_1$ in terms of known (from diagrams) functions of $\mu_0$:
\beq
   0 = \left[ \frac{\partial\Omega_1}{\partial\mu} \right]_{\mu=\mu_0}
   + \mu_1 \left[ \frac{\partial^2\Omega_0}{\partial\mu^2} \right]_{\mu=\mu_0}
        \ \Longrightarrow\
         \mu_1 = -\frac{[\partial\Omega_1/\partial\mu]_{\mu=\mu_0}}
          {[\partial^2\Omega_0/\partial\mu^2]_{\mu=\mu_0}}
       \ .
       \label{eq:mu1eq}
\eeq
At second order, we can isolate and solve for $\mu_2$, eliminating $\mu_1$
using (\ref{eq:mu1eq}).
This pattern continues to all orders: $\mu_i$ is determined by functions
of $\mu_0$ only.

Now we apply the inversion to $F = \Omega + \mu N$:
\bea
  F(N) &=& \underbrace{\Omega_0(\mu_0) + \mu_0 N}_{F_0}
  \null + \null \underbrace{
      \mu_1 \left[\frac{\partial\Omega_0}{\partial\mu}\right]_{\mu=\mu_0}
      + \Omega_1(\mu_0) + \mu_1 N
      }_{F_1}
   \nonumber \\ & & 
  \null + \underbrace{
     \mu_2 \left[\frac{\partial\Omega_0}{\partial\mu}\right]_{\mu=\mu_0}
     + \mu_1 \left[\frac{\partial\Omega_1}{\partial\mu}\right]_{\mu=\mu_0}
      + \frac{1}{2}\mu_1^2 
      \left[\frac{\partial^2\Omega_0}{\partial\mu^2}\right]_{\mu=\mu_0}
      + \Omega_2(\mu_0) + \mu_2 N       
  }_{F_2} 
  \null + \cdots\ \
  \label{eq:Fexp2}
\eea    
The $\mu_i N$ term always cancels with $\mu_i
[\partial\Omega_0/\partial\mu]_{\mu=\mu_0}$ in $F_i$ 
 for $i\ge 1$
because of Eq.~(\ref{eq:Neq}), leaving
\beq
 F(N) = F_0(N) + \underbrace{\Omega_1(\mu_0)}_{F_1} +
   \underbrace{
   \Omega_2(\mu_0) - \frac{1}{2}
   \frac{[\partial\Omega_1/\partial\mu]^2_{\mu=\mu_0}}
        {[\partial^2\Omega_0/\partial\mu^2]_{\mu=\mu_0}}
   }_{F_2}
   + \cdots \ ,
   \label{eq:Fexp3}
\eeq
where we've also used Eq.~(\ref{eq:mu1eq}) to simplify $F_2$.
The expansion for $F$ can be extended systematically, but this is all we need
here.
(Higher orders can be found by following the prescription in
Refs.~\cite{OKUMURA96,YOKOJIMA95}.)

\begin{figure}[t]
\begin{center}
  \includegraphics*[width=12.cm,angle=0]{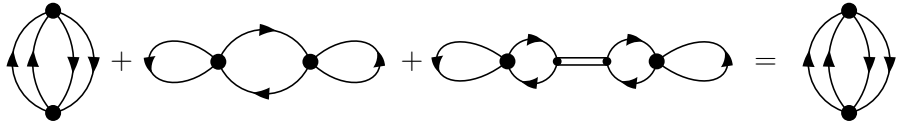}
\end{center}
\vspace*{-.2in}
\caption{Cancellation of the anomalous diagram at NLO. The double
lines represents the inverse of 
$[\partial^2\Omega_0/\partial\mu^2]_{\mu=\mu_0}$ in Eq.~(\ref{eq:Fexp3}).}
\label{fig_F2}
\end{figure}

This construction is rather general.
The Kohn-Luttinger-Ward theorem explores a particular case, the
$T\rightarrow 0$ limit, in which
the second term in $F_2$ in Eq.~(\ref{eq:Fexp3}) cancels precisely
against the anomalous diagram in $\Omega_2$, as illustrated
in Fig.~\ref{fig_F2}.
This cancellation of derivative terms and anomalous diagrams occurs to
all orders in the expansion.
(An analogous cancellation was noted in the Kohn-Sham density
functional (DFT) inversion in Ref.~\cite{PUGLIA03} and appears
again in Sec.~\ref{subsec:inversion}.)
The end result is an expression for the free-energy $F(N)$ in terms of
the diagrams used for $\Omega_i(\mu)$, only evaluated with
$\mu=\mu_0$ and excluding the anomalous diagrams (both of which simplify
the evaluation of $F(N)$!).
This is precisely the formalism used in Ref.~\cite{HAMMER00} for a
uniform
low-density Fermi gas at zero temperature, where
$\mu_0$ appeared as the Fermi energy of Eq.~(\ref{eq:mu0}).

If we think of $\mu$ as an external (constant) source coupled to the fermion
number, we can also imagine having sources coupled to other quantities
and making analogous inversions.
The generalization to multiple chemical potentials (i.e., constant
sources coupled to
conserved charges) is straightforward but adds nothing new.
There are, however, two more interesting generalizations:
\begin{enumerate}
 \item Add a second ``chemical potential'' $j$ coupled to
  a \emph{non-conserved} charge, with $j$  equal to zero in the ground
  state.  The same inversion method holds, but since we don't have
  a conserved charge such as $N$, we can't determine $j_0$ from the
  analog to Eq.~(\ref{eq:Neq}).
  Instead, we use 
  $j_0 + j_1 + j_2 + \cdots = 0$ in the ground state to solve for $j_0$
  iteratively.  
  (That is, guess $j_0$, calculate $j_i$ for $i>0$ in terms of $j_0$,
  find a new $j_0$ from $j_0 = -(j_1 + j_2 + \cdots)$, and repeat to
  self-consistency.)
  It is this method that we use here to incorporate
  pairing for a uniform system with no external fields.  
 \item Add a spatially dependent (classical)
   source $J(\xvec)$ coupled to the density operator
   $\psi^\dagger\psi$, which has expectation value 
   $\rho(\xvec) \equiv \langle \psi^\dagger\psi\rangle$.
   Then construct the functional Legendre transformation
   $F[\rho] = \Omega[J] - \int\!\rho J$, where
   $J(\xvec) 
         = \delta F[\rho]/\delta \rho(\xvec) \rightarrow 0$
         in the ground state.
   This leads to the Kohn-Sham DFT discussed in Ref.~\cite{PUGLIA03},
   where $F[\rho]$ is proportional to the effective action
   $\Gamma[\rho]$ for the composite operator $\psi^\dagger\psi$.
\end{enumerate}
In Ref.~\cite{FURNSTAHL04}, a spatially dependent 
source coupled to the kinetic energy
density is included; in this case the inversion method
leads to a generalization of conventional Kohn-Sham DFT
that is analogous to the Skyrme-Hartree-Fock energy functional
\cite{BENDER2003}. 
A natural generalization of Skyrme-Hartree-Fock-Bogoliubov
will follow by extending the present discussion to $J(\xvec)$ in finite
systems.

\subsection{Renormalization}
\label{subsect:renorm}

Starting with $W_2[\mu]$, the diagrams in Fig.~\ref{fig_Wexpansion}
may have ultraviolet divergences.
In Ref.~\cite{HAMMER00}, these divergences were regulated and
renormalized using dimensional regularization with minimal subtraction
(DR/MS).
This choice provided a clean factorization of each diagram,
with the dependence on the effective range parameters solely
in the coefficients and the
integrals depending on the Fermi momentum $\kf$ times universal
geometric factors.
The result is a very systematic power counting, with each diagram
contributing to exactly one order in the EFT expansion.
  
However, while DR/MS is convenient and efficient and correct, 
it can obscure the
renormalization process.  
In the present case, since there are subtle questions about renormalizing
effective actions with composite operators,
we adopt the power divergence subtraction (PDS) scheme of Kaplan, Savage,
and Wise \cite{KSW},
which is a generalization of DR/MS that was introduced to provide a
consistent power counting in the two-body problem when the scattering
length $a_s$ is
unnaturally large \cite{KSW,vanKolck:1998bw}.  
(An alternative for that problem is a momentum subtraction
scheme proposed by Gegelia \cite{Gegelia}.)
A parameter is introduced, which we call $\Lambda$ to avoid confusion with
the chemical potential
(this same parameter is called $\mu$ in Ref.~\cite{KSW}), as part of the
finite subtraction.
To apply PDS for natural scattering lengths, 
we take $\Lambda \ll 1/a_s$ and expand consistently
for small $\Lambda$. 
The DR/MS scheme is recovered by taking $\Lambda = 0$.
The point of keeping $\Lambda$ explicit is that
we can verify that we have successfully renormalized (to a given
order) by observing the cancellation of $\Lambda$ dependence between the
couplings and loop integrals.

For example, consider Fig.~\ref{fig_Wexpansion}(b) evaluated with
$\mu_0$.
After applying zero temperature Feynman rules, carrying out the
frequency integrals, and transforming to center-of-mass
variables, we obtain (see Ref.~\cite{HAMMER00} for further details):
\bea
  \edens_2 &=& 4 C_0^2 M \nu(\nu-1) 
       \int\! \frac{d^3 s}{(2\pi)^3} \int\! \frac{d^3 t}{(2\pi)^3}
       \left(\frac{\Lambda}{2}\right)^{3-D} 
       \int\! \frac{d^D u}{(2\pi)^D} \,\theta(\kf-|\vec{s}+\vec{t}|)
        \theta(\kf-|\vec{s}-\vec{t}|)   \nonumber \\
     & & \qquad\qquad \null \times
         [1-\theta(\kf-|\vec{s}+\vec{u}|)][1-\theta(\kf-|\vec{s}-\vec{u}|)]
           \, \frac{1}{t^2-u^2+i\epsilon}
     \ ,
   \label{energy2b}
\eea
with $\nu=2$ the spin degeneracy.
The integral over ${\bf u}$ in Eq.~(\ref{energy2b}), which is regulated
in $D$ spatial dimensions, contains a linearly
divergent term for $D=3$ coming from the piece
of the integrand without any $\theta(\kf-|\vec{s}\pm\vec{u}|)$ factors.
In PDS, with the conventions of Ref.~\cite{KSW}, this integral is:
\beq
  \left(\frac{\Lambda}{2}\right)^{3-D} \int\! \frac{d^D u}{(2\pi)^D}
    \, \frac{1}{t^2-u^2+i\epsilon}
    \stackrel{{\rm PDS}}{\longrightarrow} -\frac{1}{4\pi} (\Lambda + it)
    \ .
    \label{eq:pdsintegral}
\eeq
The imaginary part cancels with an equal and opposite imaginary part
from the remainder of the (finite) integral over $u$ [identified using
$1/(t^2-u^2+ i\epsilon)  = {\cal P}\{1/(t^2-u^2)\} - i\pi \delta(t^2-u^2)$],
so that the energy density is real.
The real parts of the integrals that contain 
$\theta(\kf-|\vec{s}\pm\vec{u}|)$ 
factors reproduce the result for $\edens_2$ given in Ref.~\cite{HAMMER00}.
The contribution proportional to $\Lambda$ is easily evaluated after
changing variables to $\vec{q} = \vec{s} + \vec{t}$ and 
$\vec{q'} = \vec{s} - \vec{t}$:
\bea
  \delta\edens_2 &=& -4 C_0^2 M \nu(\nu-1) 
       \int\! \frac{d^3 s}{(2\pi)^3} \int\! \frac{d^3 t}{(2\pi)^3} 
        \,\theta(\kf-|\vec{s}+\vec{t}|)
        \theta(\kf-|\vec{s}-\vec{t}|) 
        \frac{1}{4\pi} \Lambda  \nonumber \\
  &=& 
   -\frac{1}{8\pi} C_0^2 M (1 - 1/\nu) \Lambda \left[\nu \int^{\kf}
   \!\frac{d^3q}{(2\pi)^3}\right]^2
   = -\frac{1}{8\pi} C_0^2 M (1 - 1/\nu) \Lambda \rho^2 \ ,
   \label{eq:deltaedens}
\eea
where $\rho$ is the fermion density.
Thus, we have an apparent dependence in the energy on the arbitrary
parameter $\Lambda$.

\begin{figure}[t]
\begin{center}
  \includegraphics*[width=5.cm,angle=0]{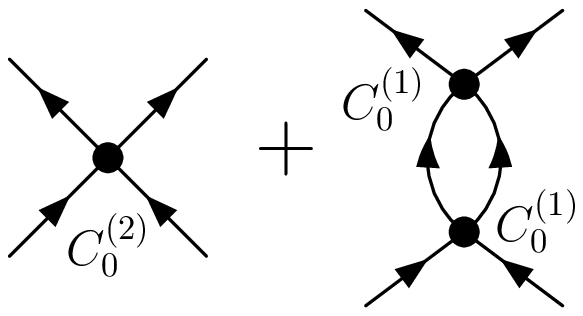}
  \raisebox{.5in}{\mbox{$\quad \Longrightarrow \quad$}}
  \includegraphics*[width=6.cm,angle=0]{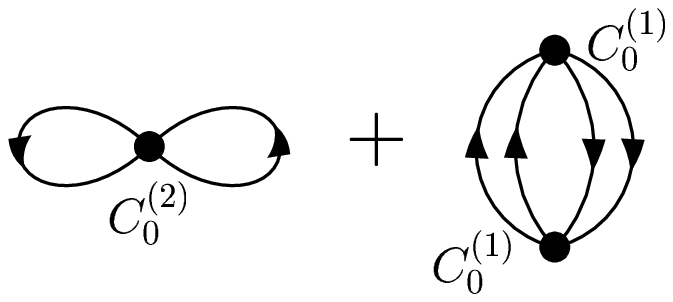}
\end{center}
\vspace*{-.2in}
\caption{Renormalization of the bubble in free space determines
$C_0^{(2)}$ to remove the linear $\Lambda$ dependence.
This implies the renormalization of the beachball  
(or ABA basketball \cite{BRAATEN}) diagram
Fig.~\ref{fig_Wexpansion}(b) at ${\cal O}(\Lambda)$.}
\label{fig:beachballrenorm}
\end{figure}

However, we also have a contribution at ${\cal O}(\Lambda)$ from the
diagram in Fig.~\ref{fig_Wexpansion}(a) from the dependence of
$C_0$ on $\Lambda$   
in the PDS scheme.
In particular, the requirement of $\Lambda$ independence for
the free-space scattering problem yields \cite{KSW}:
\beq
   C_0(\Lambda) = \frac{4\pi}{M}
     \left(
       \frac{1}{-\Lambda + 1/a_s}
     \right)
     = \frac{4\pi a_s}{M} + \frac{4\pi a_s^2}{M}\Lambda + 
     {\cal O}(\Lambda^2)
     \equiv C_0^{(1)} + C_0^{(2)} + \cdots
    \ .
    \label{eq:COPDSex}
\eeq
If we substitute the leading (independent of $\Lambda$)
term in Eq.~(\ref{eq:COPDSex}) for $C_0$
in Eq.~(\ref{eq:deltaedens}), we get a term proportional to
$\kf^6 a_s^2\Lambda$.  This cancels precisely with the contribution for
the Hartree-Fock diagram Fig.~\ref{fig_Wexpansion}(a) evaluated with 
a $C_0^{(2)}$ vertex (as implied by Fig.~\ref{fig:beachballrenorm}),
which removes the linear $\Lambda$ dependence.
Higher-order insertions of $C_0$ cancel between these diagrams and
higher-order diagrams in the $\kf$ expansion.


\section{Effective Action and Inversion Method with Pairing}
\label{sect:effact}

\subsection{Generalized Generating Functional and Inversion}
\label{subsec:inversion}

To allow the probing of pairing correlations,
we extend our generating functional to include the source $j(x)$ coupled
to the pair density,
\beq
    Z[\mu,j] = e^{-W[\mu,j]}
      = \int\! D\psi_\alpha D\psi_\alpha^\dag 
      \ e^{-\int\! d^4x\ \{{\cal L}_E\,-\,   \mu
    \psi_\alpha^\dag(x) \psi_\alpha(x)
    \,+\, j[\psidagup(x)\psidagdown(x)+ \psidown(x)\psiup(x)]
    + \frac12\zeta j^2\}}
    \ .
   \label{partfunca}
\eeq
In general we should include both $j(x)$ and $j^\ast(x)$, but in the
present case it suffices to consider real and constant $j$.
The Grassmann sources $\eta$ and $\eta^\dagger$ are not shown, but
they can be added to generate a perturbative (i.e., diagrammatic)
expansion for $Z$ or $W$.

\begin{figure}[t]
\begin{center}
  \includegraphics*[width=10.cm,angle=0]{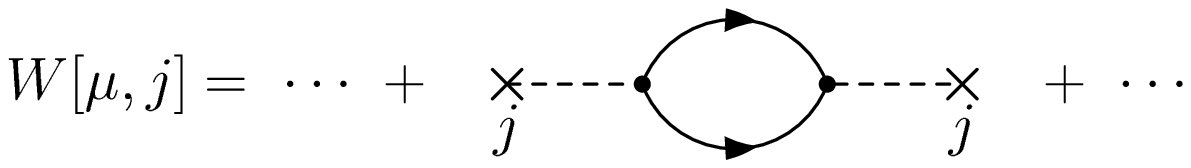}
\end{center}
\caption{Feynman diagram at second order in a perturbative expansion
of $W[\mu,j]$ in $j(x)$.}
\label{fig:Wdivergence}
\end{figure}

As discussed in Ref.~\cite{ZINNJUSTIN}
(see also Verschelde et al.\ \cite{VERSCHELDE95,VERSCHELDE97,KNECT01}), 
to make finite a generating
functional with a source coupled to
a local composite operator, we will 
need to add as counterterms
the most general linear combination of vertices of non-positive dimension
made from that operator and the source.
A perturbative expansion of $W$
in powers of $j$ immediately reveals a divergence,
shown diagrammatically in Fig.~\ref{fig:Wdivergence}.
To renormalize
in the present case, we include the term $\frac12\zeta j^2$,
which plays a role in the renormalization (see below)
analogous to the quadratic charged scalar term 
in the auxiliary field approach \cite{NAGAOSA,STONE}.
We note that this term is also the source of potential problems with
the interpretation of the effective action \cite{BANKS76}; 
we comment further on this issue in Sect.~\ref{sect:Gammaone}.
Other counterterms in Eq.~(\ref{partfunca}), including one 
proportional to $j(\psidagup\psidagdown + \psidown\psiup)$,
are implicit.

The generating functional implies that for constant
$j$ and a uniform system,
\bea
  \rho &\equiv& \langle \psidagger\psi \rangle_{\mu,j}
  = -\frac{1}{\beta V}
  \frac{\partial W[\mu,j]}{\partial \mu}
   \ , \label{eq:rhoexact} 
   \\
  \phi &\equiv& \langle \psidagup\psidagdown + \psidown\psiup
      + \zeta j \rangle_{\mu,j}
  = \frac{1}{\beta V}
  \frac{\partial W[\mu,j]}{\partial j}
   \ .  \label{eq:phiexact}
\eea
Note that $\phi$ corresponds to the usual definition in the 
ground state, where $j=0$. 
The explicit dependence of $\phi$ on $j$ through $\zeta$
is a consequence of the need to renormalize
$\phi$ \cite{VERSCHELDE95,VERSCHELDE97,KNECT01}.
We generalize the KLW derivation from Sect.~\ref{subsect:subsec} by carrying 
out a double Legendre transformation,  making the associations
\beq
  F \rightarrow \frac{1}{\beta}\,\Gamma
  \quad \mbox{and}\quad
   \Omega \rightarrow \frac{1}{\beta}\, W
   \,,
\eeq
 to switch variables to $\rho$ and $\phi$,
\beq
   \frac{1}{\beta V} \Gamma[\rho,\phi] = \frac{1}{\beta V} W[\mu,j]
       + \mu \rho   - j \phi 
     \ .
    \label{eq:fullLeg}
\eeq
We proceed in analogy to Eqs.~(\ref{eq:Omegaexp})--(\ref{eq:Fexp})
to construct the effective action $\Gamma[\rho,\phi]$ using the
inversion method.
In particular, we now have the expansions
\bea
  W[\mu,j] &=& W_0[\mu,j] + W_1[\mu,j] + W_2[\mu,j] + \cdots \ ,
    \label{eq:Wexpand} \\
  \mu &=& \mu_0 + \mu_1 + \mu_2 + \cdots \ ,
    \label{eq:muexpand} \\
  j &=& j_0 + j_1 + j_2 + \cdots \ ,
    \label{eq:jexpand} \\
  \Gamma[\mu,j] &=& \Gamma_0[\mu,j] + \Gamma_1[\mu,j] + \Gamma_2[\mu,j] 
     + \cdots \ .
    \label{eq:Gammaexpand} 
\eea

As before, the inversion is carried out by matching 
Eqs.~(\ref{eq:rhoexact}), (\ref{eq:phiexact})  and (\ref{eq:fullLeg})
order by order using (\ref{eq:Wexpand})--(\ref{eq:Gammaexpand}),
with both $\rho$ and $\phi$ 
counted formally as zeroth order.
A complication in carrying this out
is that $\phi$ will in general be renormalized order by
order, which implies contributions to the left side of
Eq.~(\ref{eq:phiexact}) beyond zeroth order.  
To avoid this, we will use the freedom
in the renormalization conditions to require that
there are no corrections to $\phi$, i.e., that $\phi$ and $\rho$
are given exactly by
the zeroth-order equations,
\beq
  \rho  = -\frac{1}{\beta V}
  \frac{\partial W_0[\mu_0,j_0]}{\partial \mu_0}
  \ ,
  \qquad
  \phi  = \frac{1}{\beta V}
  \frac{\partial W_0[\mu_0,j_0]}{\partial j_0}
   \ ,  \label{eq:zeroth}
\eeq
where it is implied that $\mu_0$ and $j_0$ are determined as functions
of $\rho$ and $\phi$.
The contributions to $\Gamma_i$ from $\mu_i\rho$ and $j_i\phi$ when
$i\geq 1$ cancel as in Eqs.~(\ref{eq:Fexp2}) and (\ref{eq:Fexp3}) 
after using Eq.~(\ref{eq:zeroth}).
The anomalous diagrams in $W_2$  cancel as in Eq.~(\ref{eq:Fexp3}) 
and Fig.~\ref{fig_F2} against additional
terms from the inversion \cite{PUGLIA03,FURNSTAHL04}, 
leaving (with $\overline W_2$ 
denoting the non-anomalous part)
\beq
  \Gamma[\rho,\phi] = \Gamma_0[\rho,\phi] + 
    \underbrace{W_1[\mu_0,j_0]}_{\Gamma_1}
    + \underbrace{\overline W_2[\mu_0,j_0]}_{\Gamma_2}
    + \cdots
    \ . 
    \label{eq:Gammacancel} 
\eeq
Thus, in practice we only calculate the non-anomalous diagrams
to this order.

To carry out a calculation, we truncate the expansion at a given order,
construct the $W_i$ needed in the corresponding truncated version
of Eq.~(\ref{eq:Gammacancel}), and determine the $\mu_i$ and $j_i$
for $i\geq 1$ from
\beq
  \mu_i = \frac{1}{\beta V}
  \frac{\partial \Gamma_i[\rho,\phi]}{\partial \rho}
  \ ,
  \qquad
  j_i  = -\frac{1}{\beta V}
  \frac{\partial \Gamma_i[\rho,\phi]}{\partial \phi}
   \ .  \label{eq:muiji}
\eeq
The values $\mu_0$ and $j_0$ are fixed by requiring both that a specified
$\rho$ to be reproduced by Eq.~(\ref{eq:zeroth}) \emph{and}
that $j = 0$, which means
\beq
   j_0 = -(j_1 + j_2 + \cdots)  \ .
\eeq
In general, these conditions can only be established numerically.


\begin{figure}[t]
\begin{center}
  \includegraphics*[width=14.cm,angle=0]{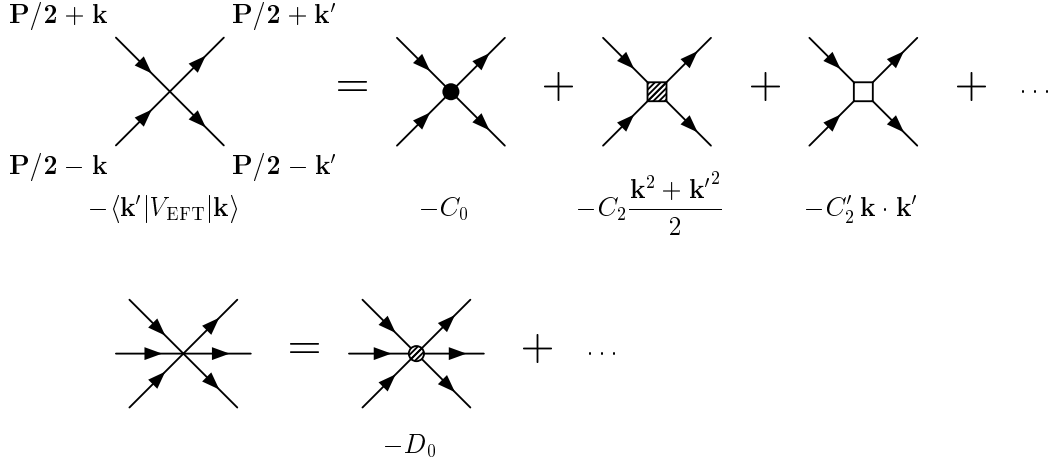}
\end{center}
\vspace*{-.2in}
\caption{Feynman rules for $-\langle\vec{k}'|V_{\rm EFT}|\vec{k}\rangle$
and the leading 3-body contact interaction.
The spin indices have been suppressed.}
\label{eftvertex}
\end{figure}

\subsection{$W_0[\mu,j]$ and $\Gamma_0[\rho,\phi]$}
\label{subsect:Wzero}

To find $W_0$ and $\Gamma_0$, we need to 
solve a non-interacting system in the presence
of external sources $\mu_0$ and $j_0$.
The path integral expression for $W_0[\mu_0,j_0]$ with constant sources is 
\beq
    Z_0[\mu_0,j_0] = e^{-W_0[\mu_0,j_0]}
      = \int\! D\psi_\alpha D\psi_\alpha^\dag 
      \ e^{-\int\! d^4x\, [\psi_\alpha^\dagger (\partial/\partial\tau  
          - \nabla^{\,2}/2M - \mu_0 )
                 \psi_\alpha\,
    +  j_0 (\psiup\psidown + \psidagdown\psidagup )
    + \frac{1}{2} \zeta^{(0)} j_0^2  ]}
    \ ,
   \label{partfunc0}
\eeq
where we restrict the $\alpha$ sum to run over spin up and spin down
and the $x$ dependence of the $\psi$ fields is implicit.
We denote the value of $\zeta$ at this order as $\zeta^{(0)}$,
which should depend on $\Lambda$ only.
Since the path integral is quadratic in the fermion fields, it
can be evaluated exactly in the Nambu-Gorkov formalism as  
$\Tr\ln(\GKSmatrix^{-1})$ (see, for example, Refs.~\cite{NAGAOSA,STONE}),
where $\GKSmatrix^{-1}$ is the quadratic part of the Lagrangian
and $\GKSmatrix$ is the associated Kohn-Sham Green's function.
Alternatively,
a canonical Bogoliubov transformation solves the problem exactly and is
particularly transparent and familiar.
(In a finite system, we would solve the zeroth-order system by
solving Hartree-Fock-Bogoliubov equations with external Kohn-Sham sources
corresponding to a spatially dependent chemical potential and a spatially
 dependent pair density.)
This is carried out in Appendix~\ref{app:canonical}
and we simply quote that result for $W_0$:
\beq
  \frac{1}{\beta V} W_0[\mu_0,j_0] = 
     \int\! \frac{d^3k}{(2\pi)^3} \, (\xi_k - E_k) 
       + \frac{1}{2} \zeta^{(0)} j_0^2
   \ ,
   \label{eq:W0expr}
\eeq
where
\beq
   \xi_k \equiv \epsk^0 - \mu_0 \ ,   
\eeq
with $\epsk^0 \equiv k^2/2M$ and
\beq
  E_k \equiv \sqrt{\xi_k^2 + j_0^2}
  \ .
  \label{eq:Ekdefa}
\eeq

The zeroth-order effective action is then
\beq
  \frac{1}{\beta V} \Gamma_0
    = \frac{1}{\beta V} W_0 + \mu_0 \rho - j_0 \phi
    \ .
    \label{eq:zerothgam}
\eeq
Equations for $\rho$ and $\phi$ follow from $W_0$:
\beq
  \rho = \frac{N}{V} = -\frac{1}{\beta V}\frac{\partial W_0}{\partial \mu_0}
     = 
      \int\! \frac{d^3k}{(2\pi)^3}\, 
      \left(
        1-\frac{\xi_k}{E_k} 
      \right)
      \ ,
    \label{eq:rhoexpr}
\eeq
and
\beq
  \phi = \frac{1}{\beta V} \frac{\partial W_0}{\partial j_0}
    = - \int\! \frac{d^3k}{(2\pi)^3}\, 
    \frac{j_0}{E_k}
    +  \zeta^{(0)} j_0
    \equiv \phibare + \zeta^{(0)} j_0
    \ .
    \label{eq:phiexpr}
\eeq
Except for the extra term proportional to $\zeta^{(0)}$,
these results are the same as from standard BCS calculations,
but with $j_0$ appearing in the
place of a constant gap.
This is illustrated further in the left panel of Fig.~\ref{fig:occupation},
which plots quantities defined in Eq.~(\ref{eq:ukvk}). 
With those standard definitions, $\rho = \frac{1}{2\pi^3}\int\! d^3k\, v_k^2$
and $\phibare = \frac{1}{2\pi^3}\int\! d^3k\, u_k v_k$.

The integral on the right side of Eq.~(\ref{eq:phiexpr}),
which we have defined as the ``bare'' pairing density $\phibare$,
is linearly divergent (see the right panel of Fig.~\ref{fig:occupation}). 
This requires regularization and then renormalization
through an appropriate choice of $\zeta^{(0)}$.
To apply the DR/PDS scheme, we have
adopted and extended the integrals derived for DR/MS in
Ref.~\cite{PaB99} in Appendix~\ref{app:pds}. 
From Eq.~(\ref{eq:Iofbeta}) we have immediately that
\beq
  \phi = - j_0\,I(0) + \zeta^{(0)}j_0 
  \ .
  \label{eq:phieqb}
\eeq
Since the interaction (i.e., $C_0$) 
should not appear explicitly in this zeroth-order equation,
requiring  $\phi$ to be the full renormalized
result independent of
$\Lambda$ uniquely determines 
\beq
  \zeta^{(0)} = \frac{\Lambda M}{2\pi} 
  \label{eq:zetazero}
\eeq
by dimensional analysis.
This results in
an analytic expression for $\phi$ in terms of Legendre functions
[see Eqs.~(\ref{eq:Idef}) and (\ref{eq:Iofbeta})]:
\beq
  \phi = - j_0\, \widetilde I(0) \ ,
  \label{eq:j0analytic}
\eeq
which is manifestly independent of $\Lambda$.
Upon applying (\ref{eq:Iofbeta}) term-by-term to the density equation,
we find (as expected) that  the $\Lambda$ dependence cancels:
\beq
   \rho = \Bigl( \frac{\Lambda}{2}\Bigr)^{3-D} 
          \int\!  \frac{d^Dk}{(2\pi)^D}\,
        \biggl(1 - \frac{\epsk^0 - \mu_0}{E_k}\biggr)
        = 0 - I(1) + \mu_0\, I(0)
	= -\widetilde I(1) + \mu_0\, \widetilde I(0)
        \ .
	\label{eq:rhoanalytic}
\eeq
By applying Eqs.~(\ref{eq:expansions}),
these expressions for $\phi$ and $\rho$ can be expanded at small
$x \equiv j_0/\mu_0$ [see Sect.~\ref{sect:Gammaone} and
Eq.~(\ref{eq:expansions})].


\subsection{Feynman Rules for $W_i$}
\label{subsect:rules}

The Feynman rules for the contributions to $W_i[\mu,j]/\beta V =
\Omega_i/V$
from the diagrams in Fig.~\ref{fig_Wexpansion}
are similar to those given in Ref.~\cite{HAMMER00}
(which in turn take the standard form,
{\it e.g.}, see Ref.~\cite{FETTER71,NEGELE88}),
but with Nambu-Gorkov matrix propagators.
These propagators arise from the diagonalization of $W_0$.
The only difference from standard treatments of 
superconductivity (e.g., see \cite{MAHAN00})
is that $j_0$ appears instead of a constant gap.

The basic Feynman rules in the zero-temperature limit
are as follows.
First, draw all distinct, fully connected diagrams contributing to 
a given order in the EFT expansion of $W$.
(In this paper we determine diagrams according to the $\kf$ expansion
for a dilute gas without pairing, so the power counting
is the same as in Ref.~\cite{HAMMER00}.) 
Distinct diagrams are those that cannot be deformed to coincide
with each other, including the direction of arrows.
To evaluate a diagram:
\begin{enumerate}
\item Assign nonrelativistic four-momenta (frequency $k_0$
and three-momentum ${\bf k}$) 
to all lines and enforce four-momentum conservation
at each vertex.
At finite temperature, the frequency is a discrete
Masubara fermion frequency \cite{FETTER71}.
\item 
For each vertex, include the corresponding term
from the two-body effective potential $-\langle\vec{k}'|V_{\rm
EFT}|\vec{k}\rangle$ (or analogous three-body interaction), 
as shown in Fig.~\ref{eftvertex}.
(Recall that ${\bf k}$ and ${\bf k'}$ are {\em relative\/} momenta.) 
The Nambu-index structure of the vertices 
has been suppressed in Fig.~\ref{eftvertex}.
For spin-independent interactions, the two-body vertices have
the structure $[(\tau_3)_{\alpha\gamma}(\tau_3)_{\beta\delta}
\pm (\tau_3)_{\alpha\delta}(\tau_3)_{\beta\gamma}]$,
where $\alpha,\beta$ are the Nambu indices of the incoming lines
and $\gamma,\delta$ are the Nambu indices of the outgoing lines.
The plus sign applies to $C_0$ and $C_2$ vertices and the minus sign
to $C'_2$ vertices.

For each internal line include the matrix propagator
$-\Gmatrix (\kt)$, where
$\kt \equiv (k_0,\vec{k})$ is the four-momentum assigned to the line,
and
\beq
  \Gmatrix(\kt) =
    \left(
     \begin{array}{cc}
       \Geucl(k_0,\vec{k}) & \Feucl(k_0,\vec{k}) \\
       \Feucl^\dagger(k_0,\vec{k}) & -\Geucl(-k_0,\vec{k})
     \end{array}
    \right)
    \equiv
    \left(
      \begin{array}{cc}
        \Geucl{}_{\kt}  &  \Feucl{}_{\kt} \\
        \Feucl^\dagger{}_{\kt} & -\wt\Geucl{}_{\kt}
      \end{array}
    \right)
\eeq
with
\beq
    \Geucl(\kt) = \frac{u_k^2}{ik_0-E_k}
      + \frac{v_k^2}{ik_0+E_k}
      = \frac{1}{2E_k}
      \left[
        \frac{E_k+\xi_k}{ik_0-E_k} +
        \frac{E_k-\xi_k}{ik_0+E_k}
      \right]
      \label{eq:Geucl}
\eeq
and
\beq
    \Feucl(\kt) = \Feucl^\dagger(\kt)
       = -u_k v_k
       \left[
        \frac{1}{ik_0-E_k} -
        \frac{1}{ik_0+E_k}
       \right]
      = 
        \frac{j_0}{2E_k}
       \left[
        \frac{1}{ik_0-E_k} -
        \frac{1}{ik_0+E_k}
       \right]
       \ .
\eeq
(Note: These conventions agree with those in Refs.~\cite{FETTER71}
and \cite{MAHAN00} and
differ by a sign from those in Ref.~\cite{NEGELE88}.)
The $u_k$'s and $v_k$'s are defined in the Appendix
in Eq.~(\ref{eq:ukvk}) and are seen here
to correspond to the residues at the poles of the noninteracting Green's
functions.

\item Perform summations on Nambu indices in the diagram (e.g.,
do the matrix multiplication). 
Include $-1$ for every closed fermion loop (i.e., for each trace).

\item Integrate over all independent momenta with a factor 
$\int\! d^4 k /(2\pi)^4$
where $d^4k\equiv dk_0\,d^3k$. 
[At finite temperature, the integral $\int dk_0/2\pi$ is
replaced by the sum $1/\beta \sum_n$ over fermion frequencies
$\omega_n$ \cite{FETTER71}.]
If a spatial integral is divergent, it is defined in $D$ spatial
dimensions and renormalized using the PDS subtraction scheme as discussed in
 Ref.~\cite{KSW}.
For lines originating and ending at the same
vertex, multiply by $\exp(\pm ik_0\eta)$ and take the limit $\eta\to 0^+$
after the contour integrals have been carried out. 
Use $\exp(ik_0\eta)$ for $\Geucl{}_{\kt}$ and $\Feucl{}_{\kt}$ 
and $\exp(-ik_0\eta)$ for $\wt \Geucl{}_{\kt}$.
This procedure
ensures that the correct term in the Green's function is picked up.

\item Multiply by a symmetry factor $-1/(S \prod_{l=2}^{l_{\rm max}} (l!)^m)$
where $S$ is the number of vertex permutations that transform the diagram into
itself, and $m$ is the number of equivalent $l$-tuples of lines. Equivalent
lines are lines that begin and end at the same vertices with the
same direction of arrows.

\end{enumerate}


\section{Renormalization to Leading Order}
\label{sect:Gammaone}

In this section, we carry out the inversion procedure to leading order
(LO), which means calculating $\Gamma_1[\rho,\phi]$ and determining
equations for $\mu_0$ and $j_0$.
We will work here in the zero-temperature limit, with frequency integrals
rather than frequency sums.
The final results are well known, although to our knowledge
this precise formalism has not been used.


Since $\Gamma_1[\rho,\phi] = W_1[\mu_0[\rho],j_0[\phi]]$ (see
Ref.~\cite{PUGLIA03}), 
we first find $W_1[\mu_0,j_0]$ by applying the
Feynman rules to the diagram in
Fig.~\ref{fig_Wexpansion}(a).
The overall symmetry factor is 1/2 (one equivalent pair of lines), so
the expression is (after cancelling signs),
\bea
   \frac{1}{\beta V} W_1[\mu_0,j_0]
    &=& \frac{1}{2} C_0^{(1)} \,
    [(\tau_3)_{\alpha\gamma}(\tau_3)_{\beta\delta}
        +(\tau_3)_{\alpha\delta}(\tau_3)_{\beta\gamma}]
   \nonumber \\ & & \qquad \null\times
    \int\! \frac{d^4k}{(2\pi)^4}
    \int\! \frac{d^4k'}{(2\pi)^4}\,
    \Gmatrix(\pt)_{\gamma\alpha} 
    \Gmatrix(\pt')_{\delta\beta}
    + \frac{1}{2}\zeta^{(1)} j_0^2 
    \nonumber \\[5pt] &=&
    \frac{1}{2} C_0^{(1)}  
      \int\! \frac{d^4k}{(2\pi)^4}
      \int\! \frac{d^4k'}{(2\pi)^4}
      \bigl[
       (\Geucl{}_{\kt} + \wt\Geucl{}_{\kt}) \, 
       (\Geucl{}_{\kt'} + \wt\Geucl{}_{\kt'})
    \nonumber \\ & & \qquad \null
       - (
        \Geucl{}_{\kt}\Geucl{}_{\kt'}
        - \Feucl^\dagger{}_{\kt}\Feucl{}_{\kt'}
        + \wt\Geucl{}_{\kt}\wt\Geucl{}_{\kt'}
        - \Feucl^\dagger{}_{\kt}\Feucl{}_{\kt'}
        )
      \bigr]
    + \frac{1}{2}\zeta^{(1)} j_0^2 
    \nonumber \\[5pt] &=&
    C_0^{(1)}  
      \int\! \frac{d^4k}{(2\pi)^4}
      \int\! \frac{d^4k'}{(2\pi)^4}
      \,\bigl[
       \Geucl{}_{\kt} \wt\Geucl{}_{\kt'} +
       \Feucl^\dagger{}_{\kt} \Feucl{}_{\kt'}
      \bigr]
    + \frac{1}{2}\zeta^{(1)} j_0^2 
    \ ,
     \label{eq:W1expr}
\eea
where we've suppressed 
the $e^{\pm ik_0\eta}$ convergence factors and
(except for $\zeta^{(1)}$) additional counterterm corrections discussed below.
The individual integrals we need are:
\bea
  \int\! \frac{d^4k}{(2\pi)^4}\, \Geucl(\pm k_0,\kvec)\,e^{\pm ik_0\eta} 
   &=&
   \int\! \frac{d^3k}{(2\pi)^3}\, v_k^2
   = \int\! \frac{d^3k}{(2\pi)^3}\, \frac{E_k-\xi_k}{2E_k}
   = \frac{1}{2}\rho
   \ ,
   \label{eq:Gtad} \\
  \int\! \frac{d^4k}{(2\pi)^4}\, \Feucl(k_0,\kvec)\,e^{ik_0\eta} 
   &=&
   \int\! \frac{d^3k}{(2\pi)^3}\, u_k v_k
   = -j_0 
   \left(\frac{\Lambda}{2}\right)^{3-D}
   \int\! \frac{d^Dk}{(2\pi)^D}\, \frac{1}{2E_k}
   \equiv \frac{1}{2}\phibare
   \ , \
   \label{eq:unrenorm}
\eea
(and $\Feucl = \Feucl^\dagger$).  The 
integral in Eq.~(\ref{eq:unrenorm}) has a linear
divergence, which we've regulated in $D$ spatial dimensions
(and identified $\phibare$ accordingly).

\begin{figure}[t]
\begin{center}
  \includegraphics*[width=15.cm,angle=0]{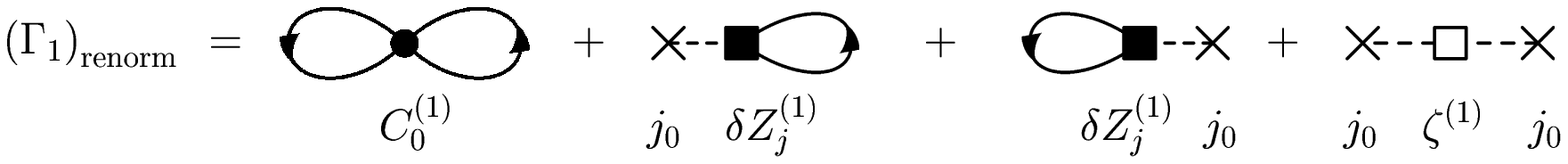}
\end{center}
\vspace*{-.2in}
\caption{Diagrammatic representation of the renormalization of
$\Gamma_1$.}
\label{fig:gam1renorm}
\end{figure}

In addition to the $\frac12\zeta^{(1)} j_0^2$ counterterm
in Eq.~(\ref{eq:W1expr}), there is a counterterm 
$\delta Z_j^{(1)}
j (\psidagup\psidagdown + \psidown\psiup)$, which contributes to $W_1$
a term proportional to $j_0 \phibare$
(see Fig.~\ref{fig:gam1renorm}).
Our renormalization
prescription  is to choose this counterterm and $\zeta^{(1)}$
so that we convert $\phibare$
to the renormalized $\phi$ from Eq.~(\ref{eq:phiexpr}) [with $\zeta^{(0)}$ 
from Eq.~(\ref{eq:zetazero})].
This takes 
$\Gamma_1 = W_1$ 
from a function of
$\rho$ and the bare $\phibare$ to a function of $\rho$
and the renormalized
$\phi$.
This prescription fixes the
counterterms 
and is analogous to that described for a 1PI
effective action in Sect.~11.4 of Ref.~\cite{PESKIN95} for tadpoles.
Since the end result is simply the replacement $\phibare \rightarrow
\phi$, there is no need to consider the explicit counterterms. 
We return below to the issue of additional finite contributions
to $\zeta$ at this order. 

This renormalization prescription allows us to write $W_1$ in
Eq.~(\ref{eq:W1expr}) directly in terms of $\rho$ and $\phi$,
yielding immediately the renormalized expression for $\Gamma_1$:
\beq
  \frac{1}{\beta V} \Gamma_1[\rho,\phi] = \frac14 C_0^{(1)}\rho^2 + \frac14
  C_0^{(1)} \phi^2 \ .
  \label{eq:Gamma1result}
\eeq
With this result, we can evaluate $\mu_1$ and $j_1$ directly:
\beq
  \mu_1 =  \frac{1}{\beta V} \frac{\partial \Gamma_1}{\partial \rho}
    = \frac12 C_0^{(1)} \rho 
    \ ,
   \label{eq:mu1expr}
\eeq
and
\beq
  j_1 = - \frac{1}{\beta V} \frac{\partial \Gamma_1}{\partial \phi}
    = -\frac12 C_0^{(1)} \phi 
    \ .
   \label{eq:j1expr}  
\eeq
More generally, we will need to eliminate $\mu_0$ and $j_0$
from the $W_i$ in
favor of $\rho$ and $\phi$ in order to construct the $\Gamma_i$.
Since the particle number (and hence the density for a uniform system)
is given, the correction $\mu_1$ is known.  
In contrast, the pairing density $\phi$, which determines $j_1$, is not
fixed by Eq.~(\ref{eq:phiexpr}) unless $j_0$ (and $\mu_0$) is known.
But since we take $j = 0$ in the ground state, and $j = j_0 + j_1$ to
this order, we have for the ground state that
\beq
  j_0 = -j_1 = \frac12 C_0^{(1)} \phi
  \ .
  \label{eq:j0LO}
\eeq
Substituting from Eq.~(\ref{eq:phiexpr}), we get the ``gap equation''
\beq
  j_0 = -\frac{C_0^{(1)}}{2}
  \left[ 
     \left(\frac{\Lambda}{2}\right)^{3-D}
   \int\! \frac{d^Dk}{(2\pi)^D}\,
     \frac{j_0}{E_k}
     - \zeta^{(0)}j_0
   \right]  
   = -\frac{C_0^{(1)}}{2}
   j_0 [ I(0) - \zeta^{(0)} ] \ ,
     \label{eq:j0gap}
\eeq
where we've used
$I(0)$ as defined in Appendix~\ref{app:pds}.
This has the trivial solution $j_0 = 0$ but also the possibility of a
non-zero ``superconducting'' solution for $j_0$ (and therefore
$\phi$) for $C_0^{(1)}<0$.

If we evaluate the gap equation 
in DR/PDS
using $I(0)$ from Eq.~(\ref{eq:Idef})
for $D=3$ and $\zeta^{(0)}$ from Eq.~(\ref{eq:zetazero}),
the $\Lambda$ dependence is precisely eliminated, leaving
\beq
  1 = 
   - \frac{C_0^{(1)}}{2} \widetilde I(0)
  = -\sqrt{2M\mu_0} |a_s| (1+x^2)^{1/4} P^0_{1/2}
    \left(\frac{-1}{\sqrt{1+x^2}}\right)
    \ ,
    \label{eq:gap1}
\eeq
with $\widetilde I(0)$ defined in Eq.~(\ref{eq:Iofbeta}).
With $\mu_0 = \kf^2/2M$, this is the same gap equation 
derived in Ref.~\cite{PaB99}, with $j_0$ taking the place of the
gap $\Delta$.
If we use the leading term in the expansion of $\widetilde I(0)$ from
Eq.~(\ref{eq:expansions}), we find
\beq
  x = \left|\frac{j_0}{\mu_0}\right| \approx \frac{8}{e^2} e^{-\pi/2\kf|a_s|}
   \ ,
   \label{eq:xapprox}
\eeq
as usual for the BCS weak-coupling limit.
This is a very good approximation even for moderate values of
$\kf |a_s|$; in particular, the approximation for $x$ is less
than the exact solution to Eq.~(\ref{eq:gap1}) by 0.06\%,
0.8\%, 2.7\%, 3.8\% for $\kf |a_s| = 0.5$, 1.0, 2.0, 3.0, respectively.
If we expand the expression for $\rho$ in Eq.~(\ref{eq:rhoanalytic}),
we find that the $\log x$ dependence from the leading term cancels, leaving
\beq
  \rho = \frac{\kf^3}{3\pi^2}
    \left[ 1 + \frac{3}{16} x^2 \left(1 - 2 \log\frac{x}{8}
      \right)
    \right]
    \ .
\eeq
Thus we get the free Fermi gas result with a correction that depends
on $x$.  For $\kf |a_s| \approx 0.5$, we find $x \approx 0.05$, and
the correction is about 0.5\%. 

The renormalized effective action to this order is found by combining
Eqs.~(\ref{eq:zerothgam}) and (\ref{eq:Gamma1result}), and using the
results in Appendix~\ref{app:pds}:  
\bea
  \frac{1}{\beta V} (\Gamma_{0}+\Gamma_{1})
   &=&
    \Bigl( \frac{\Lambda}{2}\Bigr)^{3-D} 
    \!\int\!  \frac{d^Dk}{(2\pi)^D} \, (\xi_k - E_k) 
       + \frac{1}{2} \zeta^{(0)} j_0^2 
       + \mu_0 \rho -j_0\phi
   + \frac14 C_0^{(1)}\rho^2 
    + \frac14  C_0^{(1)} \phi^2
   \nonumber    \\
   &=&
   0 -I(2) + 2 \mu_0 I(1) - (\mu_0^2 + j_0^2) I(0)
    + \frac12 \frac{\Lambda M}{2\pi} j_0^2
       + \mu_0 \rho -j_0\phi
     \nonumber \\
   & & \quad \null
   + \frac14 C_0^{(1)}\rho^2 
    + \frac14  C_0^{(1)} \phi^2 
   \nonumber    \\
   &=&
   - \widetilde I(2) + 2 \mu_0 \widetilde I(1) 
    - (\mu_0^2 + j_0^2) \widetilde I(0)
       + \mu_0 \rho -j_0\phi
   + \frac14 C_0^{(1)}\rho^2 
    + \frac14  C_0^{(1)} \phi^2 \ , 
    \label{eq:effact}    
\eea
which is manifestly independent of $\Lambda$.
To find the energy density, we evaluate 
at the stationary point: 
%
\beq
    \frac{E}{V} = \frac{1}{\beta V}\left.(\Gamma_0 + \Gamma_1)
                  \right|_{j_0 = -\frac12 |C_0^{(1)}|\phi}
      =
       \null - \widetilde I(2) + \mu_0 \widetilde I(1)  
       - \frac{|C_0^{(1)}|}{4}\rho^2
         - \frac{1}{|C_0^{(1)}|}\mu_0^2 x^2
         \label{eq:tophalf}
   \ ,
\eeq
where we've used Eqs.~(\ref{eq:gap1}) and
(\ref{eq:rhoanalytic}) to reach the final form.

In practice, however, a different renormalization 
can be more useful for the numerical
solution of the self-consistent equations and for extending the results
to inhomogeneous finite systems in the local density approximation (LDA)
and to higher orders.
The conventional approach to renormalizing the gap equation for
a delta-function interaction
is based on the observation that the unrenormalized gap equation
(with a gap $\Delta$ taking the place of $j_0$ in $E_k$), 
\beq
  1 = 
    \frac{|C_0|}{2} 
       \int \frac{d^3k}{(2\pi)^3} 
       \frac{1}{E_k}  \ ,
\eeq
has the same linear ultraviolet divergence as
the unrenormalized expression for zero-energy scattering in terms of the
scattering length \cite{MARINI98,PaB99}, 
\beq
  \frac{M |C_0|}{4\pi a_s} + 1
   = 
    \frac{|C_0|}{2} 
       \int \frac{d^3k}{(2\pi)^3} 
       \frac{1}{\epsk }
       \ .
     \label{eq:freespace}
\eeq
Thus, one can eliminate the unrenormalized $C_0$ directly 
(e.g., by subtracting the
equations and dividing out $C_0$) to obtain:
\beq
  \frac{M}{4\pi a_s} = \frac{1}{C_0^{(1)}} = - \frac{1}{2}
     \int \frac{d^3k}{(2\pi)^3}
       \left[
         \frac{1}{E_k} - \frac{1}{\epsk}
       \right] \ ,
       \label{eq:renorm}
\eeq
which is finite and 
can be evaluated using elliptic integrals \cite{MARINI98}.
This same result is obtained within the present formalism by comparing
Eqs.~(\ref{eq:phiexpr}) and (\ref{eq:renorm})
and making the association:
\beq
  \zeta^{(0)} =  \frac{\Lambda M}{2\pi} 
    \rightarrow \int\! \frac{d^3k}{(2\pi)^3} \frac{1}{\epsk}\ .
    \label{eq:zetafinal}
\eeq
This connection gives us a way to eliminate $\zeta^{(0)}$ explicitly 
in the energy and $\phi$ equations, leaving
finite expressions that can be evaluated numerically.

\begin{figure}[t]
\begin{center}
  \includegraphics*[width=10.cm,angle=0]{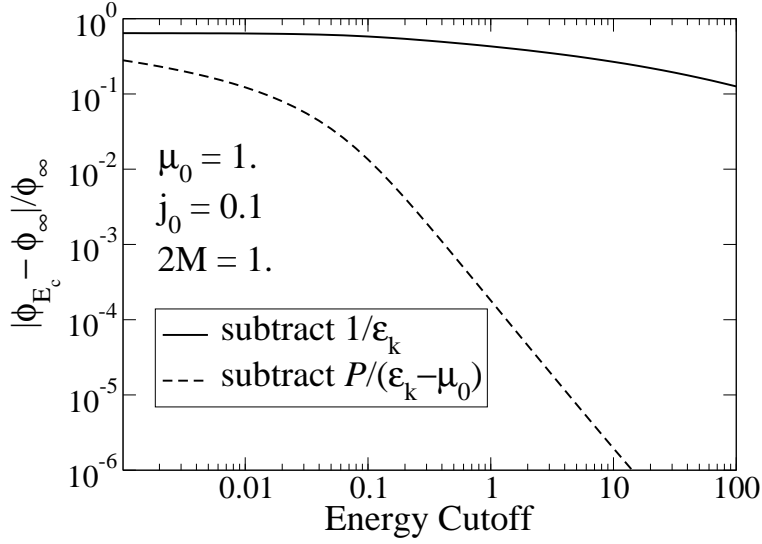}
\end{center}
\vspace*{-.2in}
\caption{Convergence of the integral for the pair density in the
  uniform system for two subtractions as a function of an energy
  cutoff $E_c$ in the integral.  The converged value of $\phi$
  is denoted $\phi_\infty$. }
\label{fig:convergence}
\end{figure}

Thus, the energy density from Eqs.~(\ref{eq:effact}) and
(\ref{eq:tophalf}) can be evaluated as
\bea
    \frac{E}{V}  &=& 
    \int\!\frac{d^3k}{(2\pi)^3}\,
       \biggl[
         \xi_k - E_k  + \frac12 \frac{j_0^2}{\epsk}
       \biggr]
       + \Bigl[\mu_0 - \frac14 |C_0^{(1)}|\rho\Bigr] \rho  
       + \frac{j_0^2}{|C_0^{(1)}|}
      \nonumber \\ 
       &=&  \int\!\frac{d^3k}{(2\pi)^3}\,
       \biggl[
         \xi_k - E_k  + \frac12 \frac{j_0^2}{E_k}
       \biggr]
       + \Bigl[\mu_0 - \frac14 |C_0^{(1)}|\rho\Bigr] \rho 
     \ .
       \label{eq:better}
\eea
The expressions for $\rho$ and $\phi$ in this approach are
  \beq
    \rho  = \int\! \frac{d^3k}{(2\pi)^3}\,
         \biggl(1 - \frac{\xi_k}{E_k}\biggr)
    \quad \mbox{and} \quad
    \phi = - j_0 \int\! \frac{d^3k}{(2\pi)^3}\,
       \left[
         \frac{1}{E_k} - \frac{1}{\epsk}
       \right]  \ .
         \label{eq:others}
  \eeq
We note that the usual normal-state LO results 
are recovered in the $j_0 \rightarrow 0$ limit:
\beq
  \frac{E}{V} \rightarrow
    \frac35 \mu_0\rho - \frac14 |C_0^{(1)}|\rho^2
    \quad \mbox{and} \quad
    \rho \rightarrow \frac{1}{3\pi^2}\kf^3
    \quad \mbox{and} \quad
    \mu_0 \rightarrow \frac{\kf^2}{2M}
    \ .
\eeq
More efficient evaluations of $\phi$ have been discussed 
by Bulgac and Yu \cite{Bulgac:2001ei,Bulgac:2001ai,Yu:2002kc}
in the context of local density approximations (LDA's) for finite systems.
In the present case, we simply illustrate how improvements can work
by comparing the evaluation of $\phi$ for two different subtractions.
We define $\phi_{E_c}$ as the evaluation of $\phi$ as a numerical
integral up to a maximum $k_c$ given by $E_c = k_c^2/2M$.  Then
we compare the two expressions for  $\phi_{E_c}$,
\beq
   \phi_{E_c}  
    = -j_0 \int^{k_c}\! \frac{d^3k}{(2\pi)^3}\, 
      \left[
        \frac{1}{E_k} - \frac{1}{\epsk}
      \right] 
    = -j_0 \int^{k_c}\! \frac{d^3k}{(2\pi)^3}\,  
      \left[
        \frac{1}{E_k} - \frac{\cal P}{\epsk-\mu_0}
      \right] 
\eeq
where the subtractions are equal in the $k_c \rightarrow \infty$ limit.
However, as shown in Fig.~\ref{fig:convergence}, the rate of convergence
is, in practice, dramatically improved by using the second subtraction.
This will be important in applications to finite systems.

The simple interpretation of the effective
action as a variational energy functional is apparently
altered by the requirement of sources appearing beyond linear
order \cite{BANKS76}.
In particular, we recall the connection between an effective
action and a variational estimate of the energy, either
in Euclidean space \cite{ZINNJUSTIN} or
in Minkowski space as a constrained minimization with the sources acting as
Lagrange multipliers \cite{WEINBERG96}.  We consider the former, with
a generic Hamiltonian $\Hhat$ and external source $J(\xvec)$ with only
spatial dependence coupled to a density  $\widehat\rho$
(which represent vectors of sources and densities in the present case):
 \beq
    \Hhat(J) = {\Hhat}  + \int\!  J\, \widehat\rho
    \label{eq:HhatJ}
 \eeq
If the ground state is isolated (and bounded from below),
then as  $\beta \rightarrow \infty$ 
the partition function in the presence of $J$, ${\cal Z}[J]$,
projects on the 
ground state of ${\Hhat}(J)$ with energy $E_0(J)$:
 \beq 
   \hspace*{-.2in}
     {\cal Z}[J] = 
     e^{-W[J]} \sim {\rm Tr\,} 
       e^{-\beta (\Hhat + J\,\widehat \rho) }
     \  \Longrightarrow\ 
   E_0(J) = \lim_{\beta\rightarrow \infty} -\frac{1}{\beta} \log 
  {\cal Z}[J]
     =  \lim_{\beta\rightarrow \infty} \frac{1}{\beta}W[J]
     \ .
 \eeq
Thus, separating out the explicit dependence on $J$ in ${\Hhat}(J)$,
 \beq
 E_0(J) = \langle {\Hhat }(J) \rangle_J 
      = \langle {\Hhat} \rangle_J + \int\! J\, 
      \langle\widehat \rho\/\rangle_{J}
      \ ,
 \eeq
where $\langle\widehat O\rangle_{J}$ means the expectation value
of $\widehat O$ in the ground state in the presence of $J$.
Combining these results, 
the expectation value of ${\Hhat}$ in the ground state
generated by $J[\rho]$ is
 \beq
   \langle {\Hhat} \rangle_J = \frac{1}{\beta}W[J] - \int J\, \rho
    = \frac{1}{\beta}\Gamma[\rho]
     \stackrel{J\rightarrow 0}{\longrightarrow}
       E_0
    \ .
 \eeq
Thus, we conclude that $\Gamma[\rho]$ provides a variational estimate of the
energy. 
However, it would seem that
this schematic argument breaks down if the source
$J$ appears nonlinearly in Eq.~(\ref{eq:HhatJ}). 
Similarly, the interpretation of $J$ as a generalized Lagrange multiplier
relies on it appearing linearly \cite{WEINBERG96}. 
Another alternative would be to remove nonlinear terms in favor
of auxiliary fields, as discussed in Refs.~\cite{VERSCHELDE95,VERSCHELDE97}.
However, it is still the case that the ground state energy is given
at a stationary point (corresponding to $J \rightarrow 0$), as long
as the operator corresponding to $\rho$ is defined through $\rho = \delta
W[J]/\delta J$.


\section{Renormalization to Next-to-Leading Order}
\label{sect:Gammatwo}

In this section, we extend the inversion procedure to next-to-leading
order (NLO) at zero temperature, 
which means calculating $\Gamma_2[\mu_0,j_0]$ and
determining equations for $\mu_0$ and $j_0$.
We start with the diagrams for $W_2$ in Fig.~\ref{fig_Wexpansion}(b) and
(c) evaluated with $C_0^{(1)}$ vertices, together with the $W_1$ diagram 
evaluated with the $C_0^{(2)}$ vertex. The latter diagram
has two terms, each of which will 
remove a
linear $\Lambda$ dependence corresponding to ultraviolet
divergences from each of the two $W_2$ diagrams.
The first term cancels a divergence and $\Lambda$ dependence 
from the beachball,
just as in Sect.~\ref{subsect:renorm}.
The second term cancels a new divergence in the anomalous diagram.
Finally, we will have the counterterm proportional to 
$j (\psidagup\psidagdown + \psidown\psiup)$, which is again chosen
to convert $\phibare$ to the renormalized $\phi$ in the
anomalous diagram Fig.~\ref{fig_Wexpansion}(c).

\begin{figure}[t]
\begin{center}
  \includegraphics*[width=15.cm,angle=0]{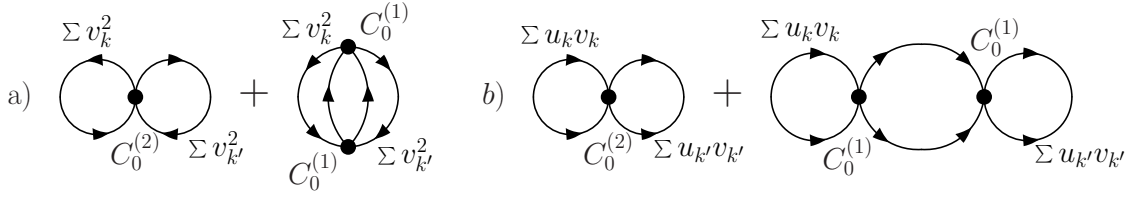}
\end{center}
\vspace*{-.2in}
\caption{Renormalization of ultraviolet divergences at NLO.  The divergent
  parts of $W_2$ evaluated with $C_0^{(1)}$ are shown with the
  corresponding $W_1$ contributions evaluated with $C_0^{(2)}$, which
  precisely cancels the $\Lambda$ dependence.}
\label{fig:W2UV}
\end{figure}

Applying the Feynman rules to $W_2^{(b)}$ and $W_2^{(c)}$ and then
carrying out the frequency integrals and simplifying through variable
transformations, 
we obtain 
\bea
  \frac{1}{\beta V} W_2^{(b)}[\mu_0,j_0] &=& 
  - \frac{1}{4} \bigl(C_0^{(1)}\bigr)^2
     \int\!\! \frac{d^4p}{(2\pi)^4}
     \int\!\! \frac{d^4k}{(2\pi)^4}
     \int\!\! \frac{d^4q}{(2\pi)^4}
     \nonumber \\ & & \qquad \null\times
     \Bigl\{
       \Tr [\Gmatrix(\wt p)\tau_3 \Gmatrix(\wt p - \wt q) \tau_3] \cdot  
       \Tr [\Gmatrix(\wt k)\tau_3 \Gmatrix(\wt k + \wt q) \tau_3]
   \nonumber \\
   & & \qquad \qquad\null -
       \Tr [\Gmatrix(\wt p)\tau_3 \Gmatrix(\wt p - \wt q) \tau_3
            \Gmatrix(\wt k)\tau_3 \Gmatrix(\wt k + \wt q) \tau_3]  
     \Bigr\}
  \nonumber \\[5pt] 
   &=&
  - \frac{1}{2} \bigl(C_0^{(1)}\bigr)^2
     \int\!\! \frac{d^4p}{(2\pi)^4}
     \int\!\! \frac{d^4k}{(2\pi)^4}
     \int\!\! \frac{d^4q}{(2\pi)^4}
     \, \bigl[ \, 
     \Geucl{}_{\wt p}\, \Geucl{}_{\wt p - \wt q}\, \Geucl{}_{\wt k}\, 
         \Geucl{}_{\wt k + \wt q}  
  \nonumber \\ & & \qquad\qquad \null +
     \Feucl{}_{\wt p}\, \Feucl{}_{\wt p - \wt q}\, \Feucl{}_{\wt k}\, 
         \Feucl{}_{\wt k + \wt q}  
      + 2\,
     \Geucl{}_{\wt p}\, \wt\Geucl{}_{\wt p - \wt q}\, \Feucl{}_{\wt k}\, 
         \Feucl{}_{\wt k + \wt q}  
     \,\bigr]
  \nonumber \\[5pt]
    &=&    
  - \bigl(C_0^{(1)}\bigr)^2
     \int\!\! \frac{d^3p}{(2\pi)^3}
     \int\!\! \frac{d^3k}{(2\pi)^3}
     \int\!\! \frac{d^3q}{(2\pi)^3}
     \frac{1}{E_p + E_k + E_{p-q} + E_{k+q}}
   \ ,
     \nonumber \\ & & \qquad\qquad \null\times
    \bigl[\,
      u_p^2\, u_k^2\, v_{p-q}^2\, v_{k+q}^2
      -2 u_p^2\, v_k^2\, (uv)_{p-q}\, (uv)_{k+q}
     \nonumber \\ & & \qquad\qquad\qquad \null
      + (uv)_p\, (uv)_k\, (uv)_{p-q}\, (uv)_{k+q}
    \,\bigr]
    \label{eq:W2b}
\eea
and
\bea
  \frac{1}{\beta V} W_2^{(c)}[\mu_0,j_0] &=& 
   \frac{1}{2} \bigl(C_0^{(1)}\bigr)^2
     \int\!\! \frac{d^4p}{(2\pi)^4}
     \int\!\! \frac{d^4k}{(2\pi)^4}
     \int\!\! \frac{d^4q}{(2\pi)^4}
     \nonumber \\ & & \qquad \null\times
     \Bigl\{
       \Tr [\Gmatrix(\wt p)\tau_3] \cdot 
       \Tr [\Gmatrix(\wt k) \tau_3\Gmatrix(\wt k)\tau_3 ] \cdot  
       \Tr [\Gmatrix(\wt q) \tau_3]
   \nonumber \\
   & & \qquad \qquad\null - 2
       \Tr [\Gmatrix(\wt p)\tau_3] \cdot 
       \Tr [\Gmatrix(\wt k) \tau_3 \Gmatrix(\wt k)\tau_3 \Gmatrix(\wt q) \tau_3]  
   \nonumber \\
   & & \qquad \qquad\null +
       \Tr [\Gmatrix(\wt p)\tau_3 \Gmatrix(\wt k) \tau_3
            \Gmatrix(\wt q)\tau_3 \Gmatrix(\wt k) \tau_3]  
     \Bigr\}
  \nonumber \\[5pt]
    &=&    
  - \bigl(C_0^{(1)}\bigr)^2
     \int\!\! \frac{d^3k}{(2\pi)^3}
   \,  \frac{1}{2 E_k}
    \bigl[
      \rho (u_k v_k)^2 + \frac12 \phibare (u_k^2 - v_k^2)
    \bigr]^2  \ .
    \label{eq:W2c}
\eea
We can identify potential ultraviolet divergences by noting that
\bea
  v_k^2 &=& \frac12 \left( 1 - \frac{\xi_k}{E_k} \right)
     \stackrel{k\rightarrow\infty}{\longrightarrow}
        \frac{j_0^2 M^2}{k^4}
    \\
  u_k^2 &=& \frac12 \left( 1 + \frac{\xi_k}{E_k} \right)
     \stackrel{k\rightarrow\infty}{\longrightarrow}
       1 - \frac{j_0^2 M^2}{k^4}
    \\
  u_k v_k &=& - \frac{j_0}{2E_k}
     \stackrel{k\rightarrow\infty}{\longrightarrow}
    - \frac{j_0 M}{k^2}
\eea
and
\beq
  \frac{1}{E_k} 
     \stackrel{k\rightarrow\infty}{\longrightarrow}
  \frac{2M}{k^2}
  \ .
  \label{eq:Eklimit}
\eeq
Using these asymptotic behaviors, we see that
in $W_2^{(b)}$ it is only the term with $u_k^2 u_p^2$ that can be
ultraviolet divergent (see Fig.~\ref{fig:W2UV}a), while in $W_2^{(c)}$ it is the term with
$u_k^4$ times $\phibare^2$ (see fig.~\ref{fig:W2UV}b).

To isolate and cancel the divergent contribution in $W_2^{(b)}$, 
it is easiest to switch variables to ${\bf s}$, ${\bf t}$, and
${\bf u}$, where
${\bf p} = {\bf s} + {\bf u}$, 
${\bf k} = {\bf s} - {\bf u}$, 
${\bf p} - {\bf q} = {\bf s} - {\bf t}$, 
${\bf k} + {\bf q} = {\bf s} + {\bf t}$, 
and we pick up a factor of 8 from the Jacobian. 
After replacing $u_{{\bf s}\pm{\bf u}}^2$ by 
$1 - v_{{\bf s}\pm{\bf u}}^2$ and keeping only the factors with $1$'s,
we find the divergent part
\beq
  \frac{1}{\beta V} [W_2^{(b)}]_{\rm divergent} =   
  -8 \bigl(C_0^{(1)}\bigr)^2
     \int\!\! \frac{d^3s}{(2\pi)^3}
     \int\!\! \frac{d^3t}{(2\pi)^3}
     \int\!\! \frac{d^3u}{(2\pi)^3}
     \frac{v_{{\bf s} - {\bf t}}^2 \,
           v_{{\bf s} + {\bf t}}^2}{E_{{\bf s}+{\bf u}} + E_{{\bf s}-{\bf u}} 
           + E_{{\bf s}+{\bf t}} + E_{{\bf s}-{\bf t}}}  
      \ .
      \label{eq:W2divergent}
\eeq
For large $u$, we see from Eq.~(\ref{eq:Eklimit}) that
Eq.~(\ref{eq:W2divergent}) has the same linear
divergence as found for ${\cal E}_2$
in the $j_0 = 0$ limit in Sect.~\ref{subsect:renorm}.
We stress that
with the divergence regulated in DR, consistent renormalization
requires that the $\Lambda$ dependence
be removed by the contribution from the Hartree-Fock diagram evaluated 
with the $C_0^{(2)}$ vertex (but now with $j_0 \neq 0$) without
adjustment.
The latter is found from the $\rho^2$ term in $W_1$ [see Eq.~(\ref{eq:W1expr})]
with $C_0^{(1)} \rightarrow C_0^{(2)} = 4\pi a_s^2\Lambda/M$,
\beq
  \frac{1}{\beta V}\delta W_2^{(b)} =  
     \frac{4\pi a_s^2\Lambda}{M}    
     \int\!\! \frac{d^3k}{(2\pi)^3}
     \int\!\! \frac{d^3k'}{(2\pi)^3} \,
     v_k^2 \, v_k'^2
     \ .
\eeq
To carry out the same sort of DR analysis as for LO, we would need to
analytically continue Eq.~(\ref{eq:W2divergent}) in $D$.  A more
practical alternative is to subtract and add the $j_0=0$ version of the
integral (also
taking $\varepsilon_s$ and $\varepsilon_t$ less than $\mu_0$), which can be
evaluated
using the integral in Eq.~(\ref{eq:pdsintegral}):
\bea
  & & -8 \bigl(C_0^{(1)}\bigr)^2
     \int\!\! \frac{d^3s}{(2\pi)^3}
     \int\!\! \frac{d^3t}{(2\pi)^3}\,
       \left(\frac{\Lambda}{2}\right)^{3-D}
     {\cal P}\!
     \int\!\! \frac{d^Du}{(2\pi)^D}
     \frac{v_{{\bf s} - {\bf t}}^2 \,
           v_{{\bf s} + {\bf t}}^2}{\varepsilon_{{\bf s}+{\bf u}} + \varepsilon_{{\bf s}-{\bf u}} 
           - \varepsilon_{{\bf s}+{\bf t}} - \varepsilon_{{\bf s}-{\bf t}}} 
      \nonumber \\
  &=& 
    -8 M \bigl(C_0^{(1)}\bigr)^2 
         \int\!\! \frac{d^3s}{(2\pi)^3}
         \int\!\! \frac{d^3t}{(2\pi)^3}\,
         v_{{\bf s} - {\bf t}}^2 \,
           v_{{\bf s} + {\bf t}}^2
       \left(\frac{\Lambda}{2}\right)^{3-D}
         \int\!\! \frac{d^Du}{(2\pi)^D}
         \frac{{\cal P}}{u^2 - t^2}
     \nonumber \\    
  &=&
    8 \frac{4\pi a_s^2\Lambda}{M}
         \int\!\! \frac{d^3s}{(2\pi)^3}
         \int\!\! \frac{d^3t}{(2\pi)^3}\,
         v_{{\bf s} - {\bf t}}^2 \,
           v_{{\bf s} + {\bf t}}^2 
      \ ,      
\eea
which, after a change of variables, is seen to equal
$-\frac{1}{\beta V}\delta W_2^{(b)}$.
Thus, the renormalized version of Eq.~(\ref{eq:W2divergent}) is obtained
with the substitution:
\beq
       \frac{1}{E_{{\bf s}+{\bf u}} + E_{{\bf s}-{\bf u}} 
           + E_{{\bf s}+{\bf t}} + E_{{\bf s}-{\bf t}}}  
   \longrightarrow
       \frac{1}{E_{{\bf s}+{\bf u}} + E_{{\bf s}-{\bf u}} 
           + E_{{\bf s}+{\bf t}} + E_{{\bf s}-{\bf t}}}
      -  
       \frac{{\cal P}}{\varepsilon_{{\bf s}+{\bf u}} + \varepsilon_{{\bf s}-{\bf u}} 
           - \varepsilon_{{\bf s}+{\bf t}} - \varepsilon_{{\bf s}-{\bf t}}} 
           \ , 
   \label{eq:substitution}
\eeq
which makes the integral explicitly finite and suitable 
for numerical evaluation.

The other part of $W_1$ with $C_0^{(2)}$ renormalizes the anomalous
$W_2$ contribution.
In particular, we isolate the divergence in Eq.~(\ref{eq:W2c}) by the
replacement $u_k^2 \rightarrow 1 - v_k^2$.  Then the divergent part
is
\beq
  \frac{1}{\beta V} [W_2^{(c)}]_{\rm divergent} =   
  - \bigl(C_0^{(1)}\bigr)^2
     \int\!\! \frac{d^3k}{(2\pi)^3}
     \frac{1}{2 E_k} \frac{1}{4}\phibare^2
     \ .
\eeq
In this case, the substitution 
\beq
       \frac{1}{E_k}  
   \longrightarrow
       \frac{1}{E_k}
      -  
       \frac{{\cal P}}{\epsk} 
\eeq
renormalizes the anomalous $W_2$.
Using Eq.~(\ref{eq:zetafinal}), we see that the additional piece 
is supplied precisely by the second term in
$W_1$ with $C_0^{(1)} \rightarrow C_0^{(2)}$, as advertised.

The prescription for deriving $\Gamma_2$ from $W_2$ was detailed
in Ref.~\cite{PUGLIA03} for the case of a source $J({\bf x})$ coupled to
the density,
and is reproduced for a constant chemical potential in
Eq.~(\ref{eq:Fexp3}).
The end result is simple: the additional terms from the inversion
method cancel precisely against the anomalous diagram (for zero-range
forces).
This same cancellation holds more generally with multiple sources and
corresponding Legendre transformations (the demonstration follows
directly by putting sources and densities into a vector, and is not shown
here).
Here one has to check that the counterterm contributions are consistent with
this cancellation.
After the removal of the anomalous diagram $W_2^{(c)}$,
the evaluation of $\Gamma_2[\rho,\phi]$ reduces to evaluating the
renormalized beachball diagram.
That is, $\Gamma_2$ is given by Eq.~(\ref{eq:W2b}) with the term
corresponding to Eq.~(\ref{eq:W2divergent}) evaluated with the substitution
in Eq.~(\ref{eq:substitution}).

While the NLO result for the energy follows formally as at LO, in practice
the solution does not follow as directly.
It particular, analytic expressions are not available for
the $\Gamma_2$ integrals and 
$j_2 = -(1/\beta V)\partial\Gamma_2/\partial\phi|_\rho$ is not simply
proportional to $\phi$ (unlike $j_1$), 
and so $j_0 = -j_1 - j_2$ does not yield
a simple gap equation.
A possible solution method is to directly solve 
$\partial[\Gamma_0 + \Gamma_1 + \Gamma_2]/\partial\phi|_\rho = 0$, 
for example via a steepest descent method,
by varying $\mu_0$ and $j_0$ while keeping $\rho$ constant using
Eq.~(\ref{eq:rhoanalytic}).
This task can be simplified by expanding about the LO solution
and/or using the expansions for small
$x = j_0/\mu_0$.  
Such numerical solutions have yet to be investigated.

\begin{figure}[t]
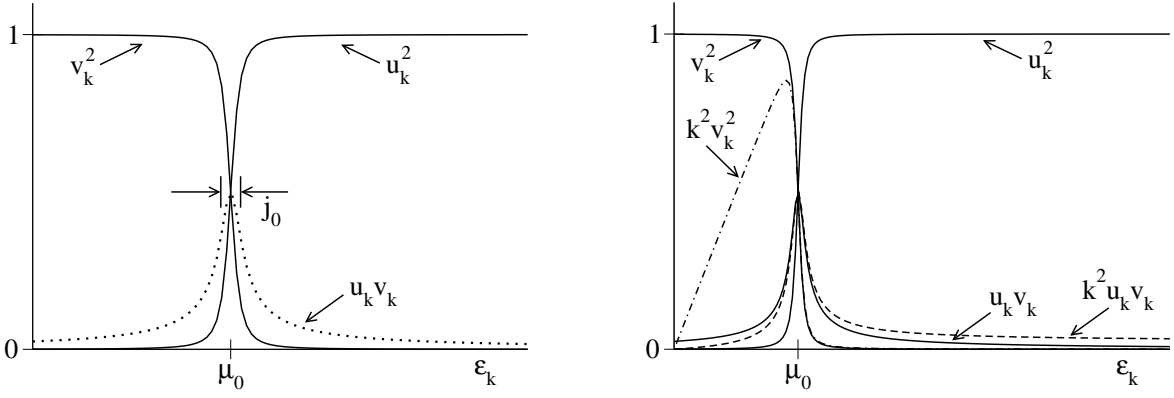

\begin{center}
  \includegraphics*[width=7.cm,angle=0]{occupation3bw}
  \hspace*{.5in}
  \includegraphics*[width=7.cm,angle=0]{occupation3_ksq_bw}
\end{center}
\vspace*{-.2in}
\caption{Left: Plot of $u_k^2$ and $v_k^2$ along with $u_k v_k$.  Note:
the width labeled ``$j_0$'' is exaggerated to make it more visible.
Right: Same plot with $k^2 v_k^2$ (dotdashes) and $k^2 u_k v_k$ added,
with an extended $x$--axis.}
\label{fig:occupation}
\end{figure}

\begin{figure}[t]
\begin{center}
  \includegraphics*[width=10.cm,angle=0]{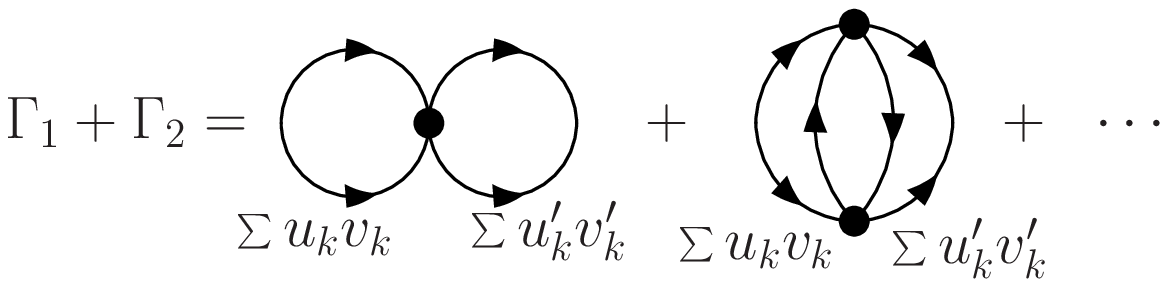}
\end{center}
\vspace*{-.2in}
\caption{Contributions to $\Gamma_1$ and $\Gamma_2$ with logarithmic
  divergences at the Fermi surface as $j_0 \rightarrow 0$.
  The vertices in each case have coefficient $C_0^{(1)}$.}
\label{fig:induced}
\end{figure}

Although the full result for $\Gamma_2[\rho,\phi]$
is only accessible numerically, we can
extract the leading correction to the gap equation analytically,
which at NLO results in a universal (i.e., independent of $a_s$)
change in the prefactor of $j_0$ (which we associate with the gap).
This contribution can be extracted by isolating the contribution to
$\Gamma_2$ that has the logarithmic divergence at the Fermi
surface of the same form as in $W_1$, which means two factors of $uv$.
We can identify this part in $W_2^{(b)}$
by letting ${\bf p} - {\bf q} \rightarrow {\bf k}$ and 
${\bf k} + {\bf q}\rightarrow {\bf k'}$ in $\Gamma_2$ to obtain
\beq
\frac{1}{\beta V}[\Gamma_2]_{\rm induced}
  = 2 \bigl(C_0^{(1)}\bigr)^2
     \int\!\! \frac{d^3k}{(2\pi)^3} u_k v_k
     \int\!\! \frac{d^3k'}{(2\pi)^3} u_{k'} v_{k'}
     \int\!\! \frac{d^3q}{(2\pi)^3}
     \frac{u^2_{{\bf k}+{\bf q}}\, v^2_{{\bf k'}-{\bf q}}}
           {{E_{{\bf k}} + E_{{\bf k'}} 
           + E_{{\bf k}+{\bf q}} + E_{{\bf k'}-{\bf q}}}}
    \ .
\eeq
We note that the terms proportional to $u_k v_k$  peak at $\mu_0$
as $j_0 \rightarrow 0$
(see Fig.~\ref{fig:occupation}). 
To extract the leading contribution from the log divergence at the Fermi
surface, we can set $j_0 = 0$ with $|{\bf k}| = |{\bf k'}| = \kf$
in the inner integral.  This means that
$E_{{\bf k}},E_{{\bf k'}} \rightarrow 0$ and the $u^2$ and
$v^2$ terms become theta functions.
As suggested by Fig.~\ref{fig:induced}, the inner integral
has the form of a non-interacting
density-density correlator or Lindhard function.
Such a correlator is defined and evaluated
in Ref.~\cite{FETTER71} for non-interacting
$T=0$ Green's functions $G_0$:
\bea
  \Pi^0({\bf q},q_0) &=&
    -2i \int\!\! \frac{d^4k}{(2\pi)^4} G^0(k) G^0(k+q) 
    \nonumber \\
   &=&
   2  \int\!\! \frac{d^3k}{(2\pi)^3} 
   \theta(|{\bf q} + {\bf k}| - \kf) \theta(\kf - |{\bf k}|)
   \nonumber \\
   & &  \null \qquad\qquad\times
   \left( 
   \frac{1}{q_0 + \epsk - \varepsilon_{{\bf q} + {\bf k}} + i\eta}
   -
   \frac{1}{q_0 + \varepsilon_{{\bf q} + {\bf k}} - \epsk - i\eta}
   \right) 
   \ .
\eea
Because of the peaking at $\mu_0$, the leading contribution is 
\bea
\frac{1}{\beta V}[\Gamma_2]_{\rm induced}
  &=& -\frac12 \bigl(C_0^{(1)}\bigr)^2
     \int\!\! \frac{d^3k}{(2\pi)^3} u_k v_k
     \int\!\! \frac{d^3k'}{(2\pi)^3} u_{k'} v_{k'}
  \left.\Pi^0({\bf k} - {\bf k'},0)
    \right|_{|{\bf k}| = |{\bf k'}|= \kf}
  \nonumber \\
  &=&
  -\frac18 \bigl(C_0^{(1)}\bigr)^2 \phibare^2
     \,
     \langle \Pi_0({\bf k}-{\bf k'}) 
     \rangle_{|{\bf k}|=|{\bf k'}|=k_{{\rm F}}}
   \ ,
\eea
where $\langle \cdots \rangle$ denotes an average over
the angle between ${\bf k}$ and ${\bf k'}$.
Using the expressions in Ref.~\cite{FETTER71} and defining
$\eta \equiv \sqrt{(\kf^2-{\bf k}\cdot{\bf k'})/2\kf^2}$,
\bea
 \langle \Pi_0({\bf k}-{\bf k'}) 
     \rangle_{|{\bf k}|=|{\bf k'}|=k_{{\rm F}}}
    &=&
      \frac{M\kf}{2\pi^2} 
        \int_0^1 d\eta\,
          \left[
          \eta + \frac{1-\eta^2}{2} \ln\left(\frac{1+\eta}{1-\eta}\right) 
          \right]  
    \nonumber \\
    &=& 
     \frac{M\kf}{2\pi^2} \frac13 (1 + 2\ln 2) \ . 
\eea

From Eq.~(\ref{eq:xapprox}),
the leading order (LO) result for $|j_0/\mu_0|$ at $T = 0$ is 
\beq
  |j_0/\mu_0|_{LO} 
     = \frac{8}{e^2}\,e^{-\pi/2\kf |a_s|}
     \ .
\eeq
After adding the renormalizations to $\Gamma_2$ that
take $\phibare$ to $\phi$,
we find that Eq.~(\ref{eq:j0LO}) becomes at NLO
\beq
   j_0 = - (j_1 + j_2) =  -\frac12 |C_0^{(1)}| \phi \left[ 1 - 
     |C_0^{(1)}|
     \frac{M\kf}{2\pi^2}\left(\frac{1 + 2\ln 2}{3}\right)\right]
  \ .  
\eeq
The final result for $|j_0/\mu_0|$ is
that the exponent is modified, which changes the prefactor, yielding
\beq
    |j_0/\mu_0|_{NLO} \approx \frac{1}{(4 e)^{1/3}}|j_0/\mu_0|_{LO}
\eeq
or, associating $j_0$ with the gap again (see below about the gap
in a Kohn-Sham DFT),
\beq
  \Delta_{NLO} \approx \frac{1}{(4 e)^{1/3}}\Delta_{LO}
  \ .
\eeq 
Thus, we recover the universal suppression of the gap in the $\kf
\rightarrow 0$ limit found in Refs.~\cite{GORKOV61,HEISELBERG00}.


\section{Summary}
\label{sect:summary}

In this paper,
we extend the effective field theory treatment of a uniform, dilute Fermi
gas to include pairing.
The starting point is 
a generating functional for the grand canonical partition function
including a chemical potential for the fermion number and a source
coupled to the pair density, which we take to be constant.
Legendre transformations with respect to the sources lead to
an effective action of the fermion and pair densities.

The effective action formalism is often used in condensed matter
applications to discuss superconductivity.
In that case, the pairing field enters as an auxiliary field.
After integrating out the fermion fields, a conventional 
one-particle-irreducible effective
action is derived by Legendre transformation with respect to the
auxiliary field.
A minimum at nonzero expectation value of this field (which is
proportional to the pairing gap) indicates spontaneously symmetry
breaking of the phase symmetry related to fermion number conservation,
which implies the normal ground state is unstable to pairing.

In contrast,
the calculation here
is carried out by adapting the inversion method proposed
in Ref.~\cite{INAGAKI92} to an EFT treatment of the dilute
Fermi gas.  
The inversion method in Ref.~\cite{INAGAKI92}
was applied to conventional BCS
superconductivity using a source
coupled to the pair creation and destruction operator.
Instead of organizing the perturbative inversion in terms of
a coupling constant (e.g., the electron charge squared), we use the EFT
expansion parameter for the natural dilute system, which is the inverse
of the resolution scale.  In the uniform system, the
ultimate dimensionless expansion parameters are products of the Fermi
momentum $\kf$ and parameters of the effective range expansion (e.g.,
the scattering length $a_s$, which is the order of the interaction range for
a natural system). 

We encounter a new renormalization problem because of ultraviolet
divergences involving the composite pair density.
To deal with it,
a new term proportional to the square of the external source $j$, with
new coupling constant $\zeta$, is added to the Lagrangian.
In our renormalization prescription, we implicitly assume that $\zeta$
has a Taylor expansion in $a_s$ to be consistent with the inversion
method. 
On dimensional grounds, a term with $\zeta$ proportional to $1/a_s$
is possible, which would be independent of $\Lambda$.  
Such a term is consistent with the development for effective actions
with local composite operators
in Refs.~\cite{VERSCHELDE95,VERSCHELDE97}, where the coefficient
of analogous terms have
a Laurent expansion starting at the inverse of the coupling.
However,
the counterterms in our prescription 
along with the source--pair-density counterterms 
suffice for us to
renormalize, with the prescription that the pair density is unchanged
from the non-interacting system.  We verify the renormalization through
NLO by using
the PDS subtraction prescription with dimensional regularization.
 
At leading order, we reproduce the results of Papenbrock and Bertsch
\cite{PaB99} for the gap and energy density.
At next-to-leading order, we reproduce the leading weak-coupling limit
derived by 
Gor'kov and Melik-Barkhudarov \cite{GORKOV61} for the modification of
the gap, and generate complete
expressions for the observables suitable for numerical
evaluation and application to finite systems in the local density
approximation (LDA).
We have focused on the zero temperature limit here, but the formalism
applies directly at finite temperature.
For example,
we can find the transition temperature by looking for 
where the non-trivial $(j_0 \neq 0)$ solution to the $j_0$
self-consistency equation disappears, which corresponds to solving for
the temperature with $j_0$ set to zero.  

The next step is to extend our results to finite systems.  This means
merging the formalism here with the density functional formalism
of Refs.~\cite{PUGLIA03,FURNSTAHL04,FURNSTAHL04b}.
The renormalization will carry over from the present discussion
via the LDA, as in Ref.~\cite{PUGLIA03}.
In practice, 
efficient subtraction schemes at leading order have been developed
by Bulgac and Yu \cite{Bulgac:2001ei,Bulgac:2001ai,Yu:2002kc}.

The extension to higher-order terms, such as $C_2$ and $C'_2$ requires further
renormalization.
In general, we expect that every term of equal or lower dimension, consistent
with symmetries, to enter.
When $C_2$ is added, for example, in the general case a new source term
proportional to $C_2 j^2 \psi^\dagger\psi$ is needed.
Extending to isospin and spin-dependent interactions is straightforward
formally,
but the renormalization issues should be reconsidered at each stage.

We have noted that the Kohn-Sham potential $j_0$ plays the same role 
at leading order 
that a gap does in a BCS treatment with a contact interaction.
The conventional definition of a gap function is an integral over
the potential and the anomalous Green's function.
For example,
Mahan defines for a local potential $V(\qvec)$ \cite{MAHAN00}:
\beq
  \Delta(\qvec) = -\frac{1}{V} \sum_{\kvec} V(\kvec)
     \Feucl(\qvec-\kvec,\tau=0)
     =  -\int\! \frac{d^4k}{(2\pi)^4}\, V(\qvec-\kvec)
            \Feucl(k_0,\kvec)\,e^{ik_0\eta}
    \ .
\eeq
Given a momentum expansion of $V$, we could define a corresponding series of
gap functions (see the appendix of Ref.~\cite{PaB99}).
However, it is important to note that the Kohn-Sham ``gaps''
are not directly connected to the single-particle spectrum.
In Ref.~\cite{FURNSTAHL04b}, the connection between the Kohn-Sham
Green's function and the full single-particle Green's function
is discussed.  There is no difference at LO, but at NLO and higher orders
the self-energies will differ, which means there is no longer
a direct correspondence between the Kohn-Sham gap function
and that extracted from the full Green's function.  
The systematics of the differences between these gaps (and also
a gap defined from even-odd staggering of ground-state energies
in finite systems)
have yet to be investigated.

The present formalism is not the only path to a nuclear DFT including
pairing.
As discussed in the introduction, an alternative to a local source coupled to
the local pair density is to use a nonlocal source and density
\cite{OLIVEIRA88,KURTH99}.
Proceeding in this way would parallel the response in particle theory
to problems uncovered by Banks and Raby, which was to focus on CJT
two-particle-irreducible effective actions.
An alternative would be to apply the auxiliary field approach in the
pairing channel.
Here a clean separation of particle-hole and particle-particle channels may be
the problem, although this is regularly a part of phenomenological approaches.
Finally,
a completely different path to DFT is outlined by Polonyi and Schwenk,
who propose a renormalization group method \cite{Schwenk:2004hm}. 
These avenues are all worth exploring.

\acknowledgments

We thank S.~Bogner,
A.~Bulgac, A.~Schwenk, and B.~Serot for useful comments
and discussions.
This work was supported in part by the National Science Foundation
under Grants No.~PHY--0098645 and No.~PHY--0354916,
by the DFG through funds provided
to the SFB/\-TR 16 \lq\lq Subnuclear structure of matter'', and by the
BMBF under contract number 06BN411.


\appendix
\section{Canonical Derivation of $W_0[\mu_0,j_0]$}
\label{app:canonical}

In this appendix, we derive an expression for $W_0[\mu_0,j_0]/\beta$,
which is proportional to the logarithm of the partition function of
the noninteracting system in the presence of $\mu_0$ and $j_0$,
in the zero temperature ($\beta\rightarrow 0$) limit.
If $\mu_0 \neq 0$ but $j_0 = 0$, we project onto 
the normal noninteracting ground state, which 
is simple to construct.  In the uniform case, we just fill the lowest
plane-wave states (which have the lowest eigenvalues) up to the Fermi
momentum $\kf$, which is related to the density by Eq.~(\ref{eq:mu0}).
This minimizes the energy for a given particle number.
When $j_0 \neq 0$, we need to construct
the state that minimizes the free energy for given chemical potential
$\mu_0$ (and given $j_0$).
Here we follow the notation and general discussion of Ref.~\cite{FETTER71} 
in using the Bogoliubov transformation, which 
finds the \emph{exact}
diagonalization in the noninteracting case with constant $j_0$.
Not surprisingly, the state that minimizes $\wh K_0 \equiv 
\wh H_0 - \mu_0 \wh N$, where $\wh H_0$ includes the
$j_0$ dependence, is a BCS-type variational state.

The non-interacting $\wh K_0$ in terms of
creation and destruction operators for the momentum states with spin up
and spin down is:
\beq
  \wh K_0 = \sum_{\kvec\alpha} (\epsk^0 - \mu_0)\,
    \akdagv{\alpha}\akv{\alpha}    
    + \sum_{\kvec} j_0\, (    
        \akdagv{\uparrow}\amkdagv{\downarrow}  
       + \amkv{\downarrow}\akv{\uparrow} ) 
       + \frac{1}{2}\zeta^{(0)}j_0^2\,V
       \ ,
       \label{eq:Kfirst}
\eeq
with $\epsk^0 = k^2/2M$.
(This definition generalizes $\wh K_0 \equiv 
\wh H_0 - \mu_0 \wh N$ from Ref.~\cite{FETTER71}.)
This result is apparent from inspection, but can be derived
formally from the field expansions for $\psi(\xvec)$ and
$\psidagger(\xvec)$.
The term proportional to $j_0^2$ is
needed to ensure $\Lambda$-independent 
results, but plays no explicit role here.
The minimized expectation value of $\wh K_0$ at fixed $V$ and $\mu$ at
$T=0$ is related to the energy by 
\beq
  \frac{1}{\beta} W_0[\mu_0,j_0]
    \stackrel{\beta\rightarrow\infty}{\longrightarrow}
     \langle \wh K_0 \rangle = \Omega_0(\mu,T=0,V)
    = (E_0 - \mu_0 N)|_{T=0} 
    \ .
    \label{eq:K0rel}
\eeq

We introduce the Bogoliubov-transformed operators $\alphak$ and
$\betak$,
\beq
  \alphak = u_k \akv{\uparrow} - v_k \amkdagv{\downarrow} \ ,
  \qquad\qquad
  \betak = u_k \amkv{\downarrow} + v_k \akdagv{\uparrow}
  \ ,
\eeq
so that
\beq
  \akv{\uparrow} = u_k \alphak + v_k \betamk^\dagger \ ,
  \qquad\qquad
  \amkv{\downarrow} = u_k \betamk - v_k \alphak^\dagger  \ .
\eeq
The anticommutation relations
\beq
 \{\alphak,\alphakp^\dagger\}  
  = \{\betak,\betakp^\dagger \} = \delta_{\kvec\kvec'} \ ,
\eeq
with all others equal to zero, imply that
\beq
   u_k^2 + v_k^2 = 1  \ .
   \label{eq:norm}
\eeq
The BCS ground state is defined so that
\beq
  \alphak\BCSket = \betak\BCSket = 0 \ .
  \label{eq:BCSdef}
\eeq
(See Ref.~\cite{FETTER71} for an explicit construction in terms of the $a$'s
and $a^\dagger$'s.)
At this point we don't know that this is actually the ground state of
the noninteracting system with non-zero $j_0$, but we do not need to
make this assumption.

By following the normal-ordering approach in Ref.~\cite{FETTER71} or
simply by applying the commutation relations, we can rewrite $\wh K_0$
with no loss of generality as:
\beq
   \wh K_0 = U + \wh H_1 + \wh H_2 \ ,
\eeq
where
\bea
  U &=& 2 \sum_{\kvec} \xi_k v_k^2 + 2j_0\sum_{\kvec} u_k v_k
       + \frac{1}{2}\zeta^{(0)}j_0^2\,V
  \ ,\\
  \wh H_1 &=& \sum_{\kvec} (\alphak^\dagger\alphak +
     \betamk^\dagger\betamk)
     [(u_k^2-v_k^2)\xi_k - 2 u_k v_k  j_0
     ]
  \ ,  \label{eq:H1}\\
  \wh H_2 &=& \sum_{\kvec}
   ( \alphak^\dagger \betamk^\dagger  + \betamk\alphak )
    [2 u_k v_k \xi_k + j_0 ( u_k^2 -  v_k^2)]
  \ .
\eea
In these expressions, 
\beq
   \xi_k = \epsk^0 - \mu_0 \ .   
\eeq
We still have freedom in our choice of $u_k$ and $v_k$, subject to the
constraint (\ref{eq:norm}).
Using this freedom, we can choose
\beq
  2 u_k v_k \xi_k = -(u_k^2 - v_k^2)j_0 \ ,
\eeq
so that $\wh H_2$ vanishes identically.
This condition is satisfied for
\beq
  u_k v_k = -\frac{j_0}{2 E_k} \,,
  \qquad
  u_k^2 = \frac12 \left( 1+ \frac{\xi_k}{E_k}\right) \,,
  \qquad
  v_k^2 = \frac12 \left( 1- \frac{\xi_k}{E_k}\right) \,,
  \label{eq:ukvk}
\eeq
with
\beq
  E_k \equiv \sqrt{\xi_k^2 + j_0^2}
  \ .
  \label{eq:Ekdef}
\eeq

Substituting from (\ref{eq:ukvk}) into Eq.~(\ref{eq:H1}), we find
that $\wh H_1$ simplifies to
\beq
  \wh H_1 = \sum_{\kvec} E_k (\alphak^\dagger\alphak +
     \betamk^\dagger\betamk)
  \ .  \label{eq:H1b}  
\eeq
Then it is clear from this equation together with 
Eq.~(\ref{eq:BCSdef}) that the
corresponding $\BCSket$ state (which is dependent on $\mu_0$ and $j_0$) 
will be the ground state of
$\wh K_0(\mu_0,j_0)$, since $U$ is a $c$-number, $E_k$ is positive
definite,
and $\langle\wh H_2\rangle \equiv 0$.

Note that there is no self-consistent gap equation at this point; for
any given value of $\mu_0$ and $j_0$, we can construct this ground state
with Eq.~(\ref{eq:ukvk}) satisfied.
The density of fermions follows immediately as
\beq
 \rho =  \frac{N}{V} = 
  \frac{1}{V} \sum_{\kvec\alpha} \BCSbra \akdagv{\alpha} \akv{\alpha} \BCSket
   = \frac{2}{V} \sum_{\kvec} v_k^2 =
   \frac{1}{V} \sum_k \left( 1 - \frac{\xi_k}{E_k} \right) \ .
\eeq
and the ground-state expectation value of $\wh K_0$, which is
equal to $U$, is
\bea
  U &=& \BCSbra \wh K_0 \BCSket =
   \left.\left( \frac{1}{\beta} W_0[\mu_0,j_0] 
      \right)\right|_{\beta\rightarrow\infty}  \nonumber \\
  &=& 2 \sum_{\kvec} \xi_k v_k^2 + 2j_0 \sum_{\kvec}u_k v_k
       + \frac{1}{2}\zeta^{(0)}j_0^2\,V \nonumber \\
   &=& \sum_{\kvec} (\xi_k - E_k)  + \frac{1}{2}\zeta^{(0)}j_0^2\,V
    \ .
\eea
Upon converting the sums to integrals and dividing by
the volume, we obtain the expression for $W_0/\beta V$ in
Eq.~(\ref{eq:W0expr}).
We emphasize that there is no approximation or truncation here.


\section{Integrals for DR/PDS}
\label{app:pds}

Here we generalize the DR/MS integrals used in Ref.~\cite{PaB99} to
the PDS subtraction scheme.
The starting point is the integral \cite{PaB99}
\beq
  \int_0^\infty \frac{\varepsilon^\alpha\, d\varepsilon}
                     {\sqrt{(\varepsilon-a)^2 + b^2}}
    = (a^2 + b^2)^{\alpha/2} 
       \left( \frac{-\pi}{\sin\pi\alpha} \right)
       P^0_\alpha (-a/\sqrt{a^2+b^2})
       \,,
      \label{eq:1stint}
\eeq
where $P^0_\alpha$ is a Legendre function.
The right side of Eq.~(\ref{eq:1stint}) can be analytically
continued in $\alpha$ to define the relevant integrals in
$D$ dimensions.
For appropriate $D$, there is an analytic formula, 
and then the answer is continued 
back to the
relevant number of spatial dimensions after the approprite subtraction
of poles (according to the prescription).

A particular integral of interest with $\alpha=0$ is 
\bea
  \left(\frac{\Lambda}{2}\right)^{3-D} 
   \int \frac{d^Dk}{(2\pi)^3} 
   \frac{1}{\sqrt{(\epsk - \mu_0)^2 + j_0^2}}
  &=&
  \left( \frac{2M}{4\pi} \right)^{D/2}
  \frac{(\Lambda/2)^{3-D}}{\Gamma(D/2)}
  \int_0^\infty \frac{\varepsilon^{(D/2-1)}\, d\varepsilon}
                     {\sqrt{(\varepsilon-\mu_0)^2 + j_0^2}}
      \nonumber \\
  &=&
  \left( \frac{2M\mu_0}{4\pi} \right)^{D/2}
  \frac{(\Lambda/2)^{3-D}}{\Gamma(D/2)}
       \left( \frac{-\pi}{\sin\pi(D/2-1)} \right)
       \nonumber \\
   & & \ \null \times
   \frac{(1+x^2)^{(D/2-1)/2}}{\mu_0} P^0_{D/2-1}
    \left(\frac{-1}{\sqrt{1+x^2}}\right)
   \,,
   \label{eq:ints}
\eea
where $x \equiv |j_0/\mu_0|$.
To apply Eq.~(\ref{eq:ints}) in the PDS scheme, we must subtract poles
in $(2-D)$.  The only such pole comes from
\beq
           \frac{-\pi}{\sin\pi(D/2-1)} 
      \longrightarrow
    \frac{2}{2-D} + \frac{\pi^2}{12}(2-D) + {\cal O}[(2-D)^2]
    \ ,
\eeq
which means the pole term is (all other factors evaluated at $D=2$):
\beq
  \left(\frac{-\Lambda}{2}\right)
  \frac{1}{\mu_0}
  \left(\frac{-2}{2-D}\right)
  \frac{2M\mu_0}{4\pi}
    = -\frac{M\Lambda}{2\pi}
       \frac{1}{(D-2)}
     \stackrel{D\rightarrow 3}{\longrightarrow} -\frac{M\Lambda}{2\pi}
     \ .
\eeq
More generally,
the basic DR/PDS integral in $D$ dimensions, 
with $x\equiv j_0/\mu_0$, is
\bea
  I(\beta) &\equiv& \Bigl( \frac{\Lambda}{2}\Bigr)^{3-D} 
    \int\!  \frac{d^Dk}{(2\pi)^D}
   \, \frac{(\epsk^0)^\beta}{E_k}
   =
   \frac{M\Lambda}{2\pi}
   \, (\mu_0)^\beta \,  \Bigl(1- \delta_{\beta,2} \frac{x^2}{2}
   \Bigr) \nonumber
   \\ & & \qquad
  \null + (-)^{\beta+1}\, \frac{M^{3/2}}{\sqrt{2}\pi} \,
     [\mu_0^2 (1+x^2)]^{(\beta+1/2)/2} \,
     P^0_{\beta+1/2}\Bigl(\frac{-1}{\sqrt{1+x^2}}\Bigr)
     \label{eq:Idef}
   \\ &\equiv&
   \frac{M\Lambda}{2\pi}
   \, (\mu_0)^\beta \,  \Bigl(1- \delta_{\beta,2} \frac{x^2}{2}
   \Bigr)
   + \widetilde I(\beta)
        \ ,
     \label{eq:Iofbeta}
\eea 
where $\widetilde I(\beta)$ is the $\Lambda$-independent part
of $I(\beta)$
 
We can expand $\widetilde I(\beta)$ with $\beta = 0,1,2$
for small $x$ as:
\bea
  \widetilde I(0)
   &=& 
     - \frac{(2 M \mu_0)^{1/2}\, M}{ \pi^2}
     \left[
       \left( 2 + \log{\frac{x}{8}} \right)
     +  \frac{1}{16}
       \left( 1 + \log{\frac{x}{8}} \right) x^2 
     + \cdots \right]
     \ ,
 \nonumber \\
  \widetilde I(1)
   &=& 
     - \frac{(2 M \mu_0)^{3/2}}{2 \pi^2}
     \left[
       \left( \frac{8}{3} + \log{\frac{x}{8}} \right)
     +  \frac{3}{16}
       \left( 1 - \log{\frac{x}{8}} \right) x^2 
     + \cdots \right]
     \ ,
  \nonumber \\
  \widetilde I(2)
   &=& 
     - \frac{(2 M \mu_0)^{3/2}\, \mu_0}{2 \pi^2}
     \left[
       \left( \frac{46}{15} + \log{\frac{x}{8}} \right)
     -  \frac{15}{16}
       \left( 1 + \log{\frac{x}{8}} \right) x^2 
     + \cdots \right]
   \ .
  \nonumber \\
  \label{eq:expansions}
\eea
%



\begin{thebibliography}{99} 

\bibitem{LEPAGE89} G. P. Lepage,
  ``What is Renormalization?'', 
        in \textit{From Actions to Answers} (TASI-89), edited by
        T. DeGrand and D. Toussaint (World Scientific, Singapore, 1989);
         ``How to Renormalize the Schr\"odinger Equation'',
         {\tt [nucl-th/9706029]}.


\bibitem{BEANE99} S.~R.~Beane, P.~F.~Bedaque, W.~C.~Haxton, D.~R.~Phillips,
                  and M.~J.~Savage,\\
                 ``From Hadrons to Nuclei: Crossing the Border'',
                  {\tt [nucl-th/0008064]}.

\bibitem{Bedaque:2002mn}
  P.~F.~Bedaque and U.~van Kolck,
  Ann.\ Rev.\ Nucl.\ Part.\ Sci.\  {\bf 52} (2002) 339 \\
  {\tt [nucl-th/0203055]}.

\bibitem{Kaplan:2005es}
  D.~B.~Kaplan,
  ``Five lectures on effective field theory,''
  {\tt nucl-th/0510023}.

\bibitem{Epelbaum:2005pn}
  E.~Epelbaum,
  Prog.\ Part.\ Nucl.\ Phys.\  {\bf 57}  (2006) 654
  {\tt [nucl-th/0509032]}.

\bibitem{Braaten:2004rn}
  E.~Braaten and H.-W.~Hammer,
  Phys.\ Rept.\  {\bf 428}  (2006) 259
  {\tt [cond-mat/0410417]}.

\bibitem{HAMMER00}
H.-W.\ Hammer and R.J.\ Furnstahl,
Nucl.\ Phys.\ A \textbf{678} (2000) 277 {\tt [nucl-th/0004043]}.
  
\bibitem{Furnstahl:2002gt}
R.J.~Furnstahl and H.-W.~Hammer,
Ann.\ Phys.\ (NY) \textbf{302} (2002) 206 {\tt [nucl-th/0208058]}.

\bibitem{PUGLIA03}S.~Puglia, A.~Bhattacharyya, and R.J.\ Furnstahl,
  Nucl.\ Phys.\ A \textbf{723} (2003) 145 {\tt [nucl-th/0212071]}.

\bibitem{FURNSTAHL04}A.~Bhattacharyya and R.J.\ Furnstahl,
  Nucl.\ Phys.\ A \textbf{747} (2005) 268 {\tt [nucl-th/0408014]}.

\bibitem{FURNSTAHL04b}A.~Bhattacharyya and R.J.\ Furnstahl,
  Phys.\ Lett.\ B \textbf{607} (2005) 259 {\tt [nucl-th/0410105]}.

\bibitem{COLEMAN88}S.~Coleman, \textit{Aspects of Symmetry\/}
   (Cambridge Univ.\ Press, 1988).

\bibitem{PESKIN95}M.E.~Peskin and D.V.~Schroeder,
 \textit{An Introduction to Quantum Field Theory} (Addison--Wesley, 1995).

\bibitem{WEINBERG96} S.~Weinberg, \textit{The Quantum Theory of
     Fields:~vol.~II, Modern Applications} (Cambridge University 
     Press, 1996).

\bibitem{NAGAOSA}N.~Nagaosa, \textit{Quantum Field Theory in Condensed Matter
     Physics} (Springer Verlag, 1999).

\bibitem{STONE}M.~Stone, \textit{The Physics of Quantum Fields} (Springer
     Verlag, 2000)
      

\bibitem{FUKUDA94}R.~Fukuda, T.~Kotani, Y.~Suzuki, and S.~Yokojima,
       Prog.\ Theor.\ Phys.\ \textbf{92} (1994) 833.
       
\bibitem{FUKUDA95}R.~Fukuda, M.~Komachiya, S.~Yokojima, Y.~Suzuki,
  K.~Okumura, and T.~Inagaki, Prog.\ Theor.\ Phys.\ Suppl.\
    \textbf{121} (1995) 1.


\bibitem{INAGAKI92}T.~Inagaki and R.~Fukuda, Phys.\ Rev.\ B \textbf{46}
     (1992) 10931.



\bibitem{VALIEV97}M.\ Valiev and G.~W.\ Fernando,
    {\tt cond-mat/9702247}.
    
\bibitem{VALIEV97b}M.\ Valiev and G.~W.\ Fernando,
    Phys.\ Lett.\ A \textbf{227} (1997) 265.

    
\bibitem{VALIEV96}M.\ Valiev and G.~W.\ Fernando,
   Phys.\ Rev.\ B \textbf{54} (1996) 7765.
   
\bibitem{CHITRA00}R.~Chitra and G.~Kotliar, Phys.\ Rev.\ B
   \textbf{62} (2000) 12715.
   
\bibitem{CHITRA01}R.~Chitra and G.~Kotliar, Phys.\ Rev.\ B
   \textbf{63} (2001) 115110.

\bibitem{Polonyi:2001uc}
J.~Polonyi and K.~Sailer,
{\tt cond-mat/0108179}.

\bibitem{KOHN60}W.~Kohn and J.~M.\ Luttinger, Phys.\ Rev.\ \textbf{118}
  (1960) 41. 
  
\bibitem{LUTTINGER60}J.~M.\ Luttinger and J.~C.\ Ward, Phys.\ Rev.\
  \textbf{118} (1960) 1417.       
   

\bibitem{OLIVEIRA88}L.~N.\ Oliveira, E.~K.~U.\ Gross, and W.~Kohn, 
  Phys.\ Rev.\ Lett.\ \textbf{60} (1988) 2430.
  
\bibitem{KURTH99}S.~Kurth, M.~Marques, M.~L\"uders, and E.~K.~U.\ Gross, 
  Phys.\ Rev.\ Lett.\ \textbf{83} (1999) 2628.

\bibitem{Stoitsov:2003pd}
  M.~V.~Stoitsov, J.~Dobaczewski, W.~Nazarewicz, S.~Pittel and D.~J.~Dean,
  Phys.\ Rev.\ C \textbf{68} (2003) 054312, and references therein.



\bibitem{COLLINS86}J.~C.\ Collins, \textit{Renormalization}
  (Cambridge Univ.\ Press, 1986).

\bibitem{BANKS76}T. Banks and S. Raby, Phys.\ Rev.\ D \textbf{14}
  (1976) 2182.
  
\bibitem{VERSCHELDE95}H. Verschelde, Phys.\ Lett.\ B \textbf{351} (1995)
242.

\bibitem{VERSCHELDE97}H. Verschelde, S. Schelstraete, and M.
Vanderkelen, Z. Phys.\ C \textbf{76} (1997) 161.

\bibitem{KNECT01}K. Knecht and H. Verschelde, Phys.\ Rev.\ D \textbf{64}
(2001) 085006.

\bibitem{MIRANSKY93}
V.~A.~Miransky,
Int.\ J.\ Mod.\ Phys.\ A \textbf{8} (1993) 135.

\bibitem{MIRANSKY97}
V.~A.~Miransky and K.~Yamawaki,
Phys.\ Rev.\ D \textbf{55} (1997) 5051
[Erratum-ibid.\ D \textbf{56} (1997) 3768].

\bibitem{MIRANSKY98}
V.~P.~Gusynin, V.~A.~Miransky and A.~V.~Shpagin,
Phys.\ Rev.\ D \textbf{58} (1998) 085023.

\bibitem{MARINI98}M.~Marini, F.~Pistolesi, and G.~C.\ Strinati, Eur.\
   Phys.\ J. B \textbf{1} (1998) 151.
  
%
\bibitem{PaB99}T. Papenbrock and G.F. Bertsch, Phys. Rev. C \textbf{59} (1999)
   2052.

\bibitem{GORKOV61}L.~P.\ Gor'kov and T.~K.\ Melik-Barkhudarov,
  Sov.\ Phys.\ JETP \textbf{13}  (1961) 1018.

\bibitem{HEISELBERG00}H.~Heiselberg, C.~J.\ Pethick, H.~Smith, and
  L.~Viverit, Phys.\ Rev.\ Lett.\ \textbf{85} (2000) 2418.


\bibitem{Furnstahl:2000we}
  R.J.~Furnstahl, H.-W.~Hammer and N.~Tirfessa,
  Nucl.\ Phys.\ A {\bf 689} (2001) 846
  {\tt [nucl-th/0010078]}.


\bibitem{NEGELE88}
J.~W.\ Negele and H.~Orland, \textit{Quantum Many-Particle Systems\/}
    (Addison-Wesley, New York, 1988).

\bibitem{FETTER71}A. L. Fetter and J. D. Walecka, \textit{Quantum Theory of
         Many-Particle Systems\/} (McGraw--Hill, New York, 1971).  

\bibitem{OKUMURA96}K.~Okumura, Int.\ J.\ Mod.\ Phys.\ A \textbf{11} (1996)
    65.      
 
\bibitem{YOKOJIMA95}S.~Yokojima, Phys.\ Rev.\ D \textbf{51} (1995) 2996. 


\bibitem{BENDER2003}
M.~Bender, P.~H.\ Heenen, and P.-G.\ Reinhard,
Rev.\ Mod.\ Phys.\ {\bf 75} (2003) 121.

  
\bibitem{KSW}
D.~B.~Kaplan, M.~J.~Savage and M.~B.~Wise,
Phys.\ Lett.\ B \textbf{424} (1998) 390
{\tt [nucl-th/9801034]};
  Nucl.\ Phys.\ B {\bf 534} (1998) 329
  {\tt [nucl-th/9802075]}.


\bibitem{vanKolck:1998bw}
  U.~van Kolck,
  Nucl.\ Phys.\ A {\bf 645} (1999) 273
  {\tt [nucl-th/9808007]}.

\bibitem{Gegelia}
J.~Gegelia,
Phys.\ Lett.\ B \textbf{429} (1998) 227;
J.\ Phys.\ G \textbf{25}  (1999) 1681
{\tt [nucl-th/9805008]}.

\bibitem{BRAATEN}E.~Braaten, private communication.


\bibitem{ZINNJUSTIN}
J. Zinn-Justin, \textit{Quantum Field Theory and Critical Phenomena}
 4th ed.\  (Oxford University Press, New York, 2002), ch.~12.

\bibitem{MAHAN00}G. D. Mahan, \textit{Many-Particle Physics\/}
         (Plenum, New York, 2000).


\bibitem{Yu:2002kc}
Y.~Yu and A.~Bulgac,
Phys.\ Rev.\ Lett.\  \textbf{90} (2003) 222501
{\tt [nucl-th/0210047]}.

\bibitem{Bulgac:2001ai}
A.~Bulgac,
Phys.\ Rev.\ C \textbf{65} (2002) 051305
{\tt [nucl-th/0108014]}.

\bibitem{Bulgac:2001ei}
A.~Bulgac and Y.~Yu,
Phys.\ Rev.\ Lett.\  \textbf{88} (2002) 042504
{\tt [nucl-th/0106062]}.

\bibitem{Schwenk:2004hm}
  A.~Schwenk and J.~Polonyi,
  {\tt nucl-th/0403011}.


\end{thebibliography}
\end{document}